%% file: PRD.tex
\documentclass[twocolumn,showpacs,aps,prd,superscriptaddress]{revtex4}
\usepackage{graphicx}
\usepackage{dcolumn}
\usepackage{amsmath}
\usepackage{epsfig}
\usepackage{subfigure}
\usepackage{verbatim}
\include{babarsym}
\def\etab     {\ensuremath{\eta_{b}(1S)}\xspace}
\def\egamma     {\ensuremath{E_{\gamma}^{*}}\xspace}

\def\figurebox#1#2#3{%
    \def\arg{#3}%
    \ifx\arg\empty
    {\hfill\vbox{\hsize#2\hrule\hbox to #2{\vrule\hfill\vbox to #1{\hsize#2\vfill}\vrule}\hrule}\hfill}%
    \else
    {\hfill\epsfbox{#3}\hfill}%
    \fi}

\newcommand{\BABARPubYear}    {11}
\newcommand{\BABARPubNumber}  {002}
\newcommand{\SLACPubNumber} {14446}
\newcommand{\LANLNumber} {1104.5254}

\begin{document}

\begin{flushleft}
\babar-PUB-\BABARPubYear/\BABARPubNumber\\
SLAC-PUB-\SLACPubNumber\\
arXiv:\LANLNumber\\[10mm]
\end{flushleft}

\title{\large \bf Study of radiative bottomonium transitions using converted photons}

\input pubboard/authors_jan2011.tex

\date{\today}

\begin{abstract}
\noindent We use $(111\pm1)$ million $\Upsilon(3S)$ and $(89\pm1)$ million $\Upsilon(2S)$ events recorded by the {\babar\ }detector at the PEP-II B-factory at SLAC to perform a study of radiative transitions between bottomonium states using photons that have been converted to \epem pairs by the detector material.
We observe $\Upsilon(3S)\rightarrow\gamma\chi_{b0,2}(1P)$ decay, make precise measurements of the branching fractions for $\chi_{b1,2}(1P,2P)\rightarrow\gamma\Upsilon(1S)$ and $\chi_{b1,2}(2P)\rightarrow\gamma\Upsilon(2S)$ decays, and search for radiative decay to the $\eta_{b}(1S)$ and $\eta_{b}(2S)$ states.
\end{abstract}

\pacs{13.20.Gd, 14.40.Pq}% PACS, the Physics and Astronomy Classification Scheme.

\maketitle

\section{INTRODUCTION}
Bottomonium spectroscopy and radiative transitions between \bbbar states can be well-described by effective potential models \cite{general_theory}.
To leading order, radiative decays are expected to be dominantly electric (E1) or magnetic (M1) dipole transitions.
In the non-relativistic limit, theoretical predictions for these decays are straightforward and well-understood. 
However, there are a few notable cases where the non-relativistic decay rates are small or zero, \emph{e.g.} in ``hindered'' M1 transitions between S-wave bottomonium such as $\Upsilon(nS)\rightarrow\gamma\eta_{b}(n'S)$ ($n>n'$), and as a consequence of small initial- and final-state wavefunction overlap in the case of $\Upsilon(3S)\rightarrow\gamma\chi_{bJ}(1P)$ decays \cite{nomenclature}; higher-order relativistic and model-dependent corrections then play a substantial role.
Measurements of these and other E1 transition rates can lead to a better understanding of the relativistic contributions to, and model dependencies of, interquark potentials.
Furthermore, because radiative transitions have a distinct photon energy signature associated with the mass difference between the relevant \bbbar states, they are useful in spectroscopic studies for mass measurements, and in the search for and identification of undiscovered resonances.

Radiative transitions within the bottomonium system have been studied previously in several experiments, such as Crystal Ball \cite{cb1, cb2}, ARGUS with converted photons \cite{argus_conv}, and iterations of CUSB \cite{cusb1, cusb2, cusb3, cusb4, cusb_chib} and CLEO \cite{cleo1, cleo2, cleo3, cleo_inclusive, cleo_new} (including an analysis of photon pair conversions in a lead radiator inserted specifically for that purpose \cite{cleo_conv}).
These analyses have focused mainly on $\chi_{bJ}(nP)$-related measurements, such as the determination of the masses and the E1 transition rates to and from $\Upsilon(mS)$ states.
More recently, the {\babar\ }experiment finished its operation by collecting large samples of data at the $\Upsilon(3S)$ and $\Upsilon(2S)$ center-of-mass (CM) energies.
These data are useful for studies of bottomonium spectroscopy and decay and have already led to the discovery of the long-sought \etab bottomonium ground state \cite{babar_etab1, babar_etab2}, an observation later confirmed by CLEO \cite{cleo_etab}.

In this paper, we present a study of radiative transitions in the bottomonium system using the inclusive converted photon energy spectrum from $\Upsilon(3S)$ and $\Upsilon(2S)$ decays.
The rate of photon conversion and the reconstruction of the resulting \epem pairs has a much lower detection efficiency than that for photons in the {\babar\ }electromagnetic calorimeter, a disadvantage offset by a substantial improvement in the photon energy resolution.
This improvement in resolution is well-suited for performing precise transition energy (hence, particle mass, and potentially width) measurements, and to disentangle overlapping photon energy lines in the inclusive photon energy spectrum.
This analysis has different techniques, data selection, and systematic uncertainties than the previous studies \cite{babar_etab1,babar_etab2,cleo_etab}, and is relatively free from complications due to overlapping transition peaks, and calorimeter energy scale and measurement uncertainties.
We report measurements of $\chi_{bJ}(2P)\rightarrow\gamma\Upsilon(2S)$, $\chi_{bJ}(1P,2P)\rightarrow\gamma\Upsilon(1S)$, observation of $\Upsilon(3S)\rightarrow\gamma\chi_{b0,2}(1P)$, and searches for the $\eta_{b}(1S,2S)$ states.

In Sec. \ref{sec_detector} we describe the {\babar\ }detector and the data samples used in this analysis.
Section \ref{sec_reco} describes the photon conversion reconstruction procedure and the event selection criteria.
Each of the following sections (Sec. \ref{sec_y3s_low} - \ref{sec_y3s_high}) individually describes the analysis of a particular region of interest in the inclusive photon energy spectrum.
Section \ref{sec_conclusion} summarizes the results obtained.
Appendix provides specific details of some systematic uncertainties related to this analysis.

\section{THE {\babar\ } DETECTOR AND DATA SAMPLES}\label{sec_detector}
The {\babar\ }detector is described in detail elsewhere \cite{babar_detector}; a brief summary is provided here.
Moving outwards from the collision axis, the detector consists of a double-sided five-layer silicon vertex tracker (SVT) for measuring decay vertices close to the interaction point, a 40-layer drift chamber (DCH) for charged-particle tracking and momentum measurement, a ring-imaging Cherenkov detector for particle identification, and a CsI(Tl) crystal electromagnetic calorimeter (EMC) for measuring the energy deposited by electrons and photons.
These detector subsystems are contained within a large solenoidal magnet which generates a 1.5-T field.
The steel magnetic flux return is instrumented with a muon detection system consisting of resistive plate chambers and limited streamer tubes \cite{babar_lst}.

The inner tracking region also contains non-instrumented support structure elements.
Interior to the SVT, the interaction region is surrounded by a water-cooled, gold-coated beryllium beam pipe.
The SVT support structure consists primarily of carbon-fiber and Kevlar$^{\textregistered}$.
The SVT, beam pipe and vacuum chamber, and the near-interaction-point magnetic elements are mounted inside a cylindrical, carbon-fiber support tube.
The inner wall of the DCH is a cylindrical tube of beryllium coated with anti-corrosion paint. 
A photon at normal incidence traverses approximately 0.01 radiation lengths ($X_{0}$) of material before reaching the SVT, and an additional 0.03$X_{0}$ before the DCH.
Due to the asymmetric energy of the incoming \epem beams, the photons in this analysis tend to be boosted in the direction of $e^{-}$ beam, increasing the typical number of radiation lengths up to 0.02$X_{0}$ and 0.08$X_{0}$ to reach the previously noted detector subsystems.
While this extra material is usually considered detrimental to detector performance, it is essential for $\gamma\rightarrow\epem$ conversions in the present analysis.

The {\babar\ }detector collected data samples of $(121\pm1)$ million $\Upsilon(3S)$ and $(98\pm1)$ million $\Upsilon(2S)$ decays \cite{semantics} produced by the PEP-II asymmetric energy \epem collider.
This corresponds to an integrated luminosity of $27.9\pm0.2$ fb$^{-1}$ ($13.6\pm0.1$ fb$^{-1}$) taken at the $\Upsilon(3S)$ ($\Upsilon(2S)$) resonance.
Approximately $10\%$ of these data (referred to here as the ``test sample'') were used for feasibility studies and event selection optimization; they are excluded in the final analysis.
The results presented in this analysis are based on data samples of $(111\pm1)$ million $\Upsilon(3S)$ and $(89\pm1)$ million $\Upsilon(2S)$ decays.
An additional $2.60\pm0.02$ ($1.42\pm0.01$) fb$^{-1}$ of data were taken at a CM energy approximately 30 \mev below the nominal $\Upsilon(3S)$ ($\Upsilon(2S)$) resonance energy, to be used for efficiency-related studies.

Large Monte Carlo (MC) datasets simulating the signal and expected background decay modes are used for the determination of efficiencies and the parameterization of lineshapes for signal extraction.
The particle production and decays are simulated using a combination of EVTGEN \cite{evtgen} and JETSET \cite{jetset}.
The radiative decays involving $\chi_{bJ}(nP)$ states are assumed to be dominantly E1 radiative transitions, and the MC events are generated with theoretically predicted helicity amplitudes \cite{helicity}.
The interactions of the decay products traversing the detector are modeled by Geant4 \cite{geant4}.

\section{EVENT RECONSTRUCTION AND SELECTION}\label{sec_reco}
\begin{figure}
\begin{center}
\subfigure{\epsfig{file={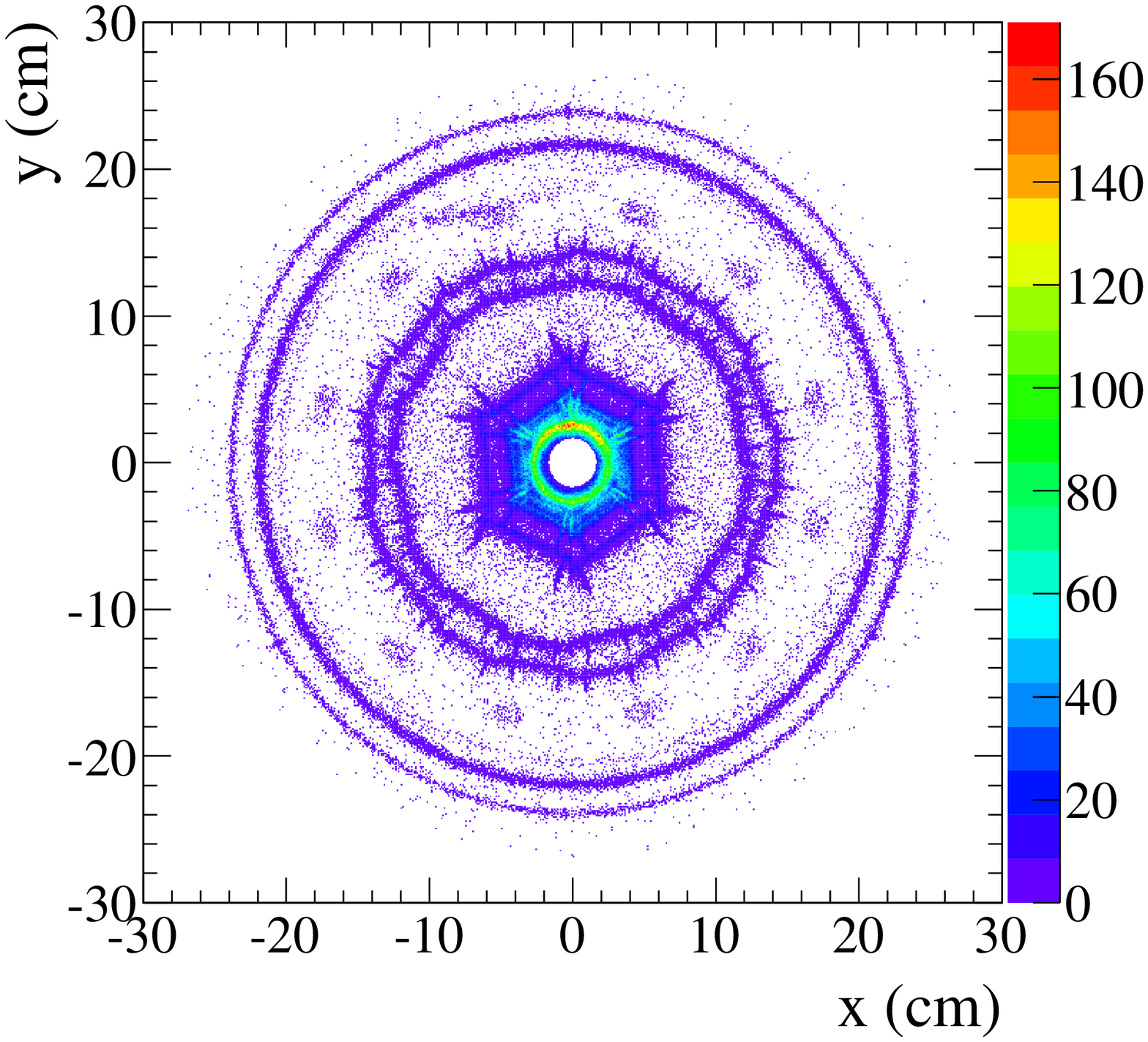},width={0.98\columnwidth},height={0.98\columnwidth}}}
\end{center}
\caption{(Color online) End view of the the {\babar\ }inner detector along the beam axis as seen by converted photons. Points indicate the number of converted photon vertices per cross-sectional area, as measured in a subset of the ``test sample'' data. From the center outwards, features of note include the beam pipe, the SVT (\emph{e.g.} hexagonal inner layers) and its support structure rods, the support tube, and the inner wall of the DCH.}
\label{fig_photograph}
\end{figure}

\begin{figure}
\begin{center}
\subfigure{\epsfig{file={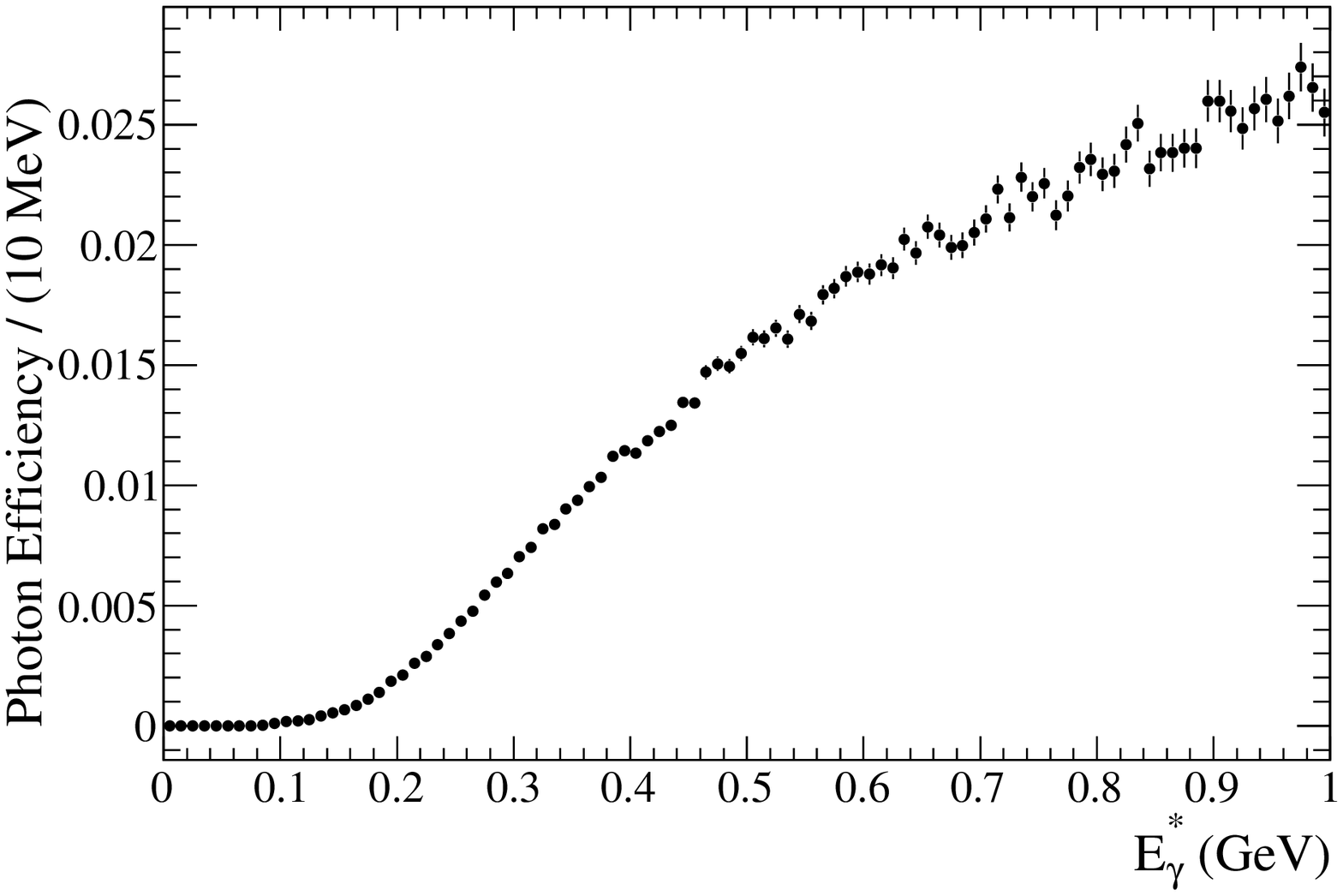},width={0.98\columnwidth},height={0.68\columnwidth}}}
\end{center}
\caption{Efficiency for the conversion and reconstruction of a photon versus photon energy as derived from a sample of generic $\Upsilon(3S)$ MC events before the optimal selection criteria have been applied.}
\label{fig_raw_eff}
\end{figure}

Photon conversions are reconstructed with a dedicated fitting algorithm that pairs oppositely charged particle tracks to form secondary vertices away from the interaction point.
The algorithm minimizes a $\chi^{2}$ value ($\chi^{2}_{fit}$) based on the difference between the measured helical track parameters and those expected for the hypothesis that the secondary vertex had originated from two nearly parallel tracks emitted from a $\gamma\rightarrow\epem$ conversion.
The $\chi^{2}_{fit}$ value includes a term to account for an observed finite opening angle between the converted tracks.
Requiring $\chi^{2}_{fit}<34$ is found to be the optimal value to select a high-purity converted photon sample.
The reconstructed converted photons are also required to have an \epem invariant mass of $m_{\epem}<30\mevcc$ (though in practice, $m_{\epem}$ is typically less than 10\mevcc).
To remove internal conversions and Dalitz decays, and to improve signal purity, the conversion vertex radius ($\rho_{\gamma}$) is required to satisfy $1.7<\rho_{\gamma}<27$ cm.
This restricts the photon conversions to the beampipe, SVT, support tube, and inner wall of the DCH, as seen in the plot of conversion vertex position for a portion of the ``test sample'' in Fig. \ref{fig_photograph}.
The efficiency for photon conversion and reconstruction versus energy in the CM frame (\egamma), as determined from a generic $\Upsilon(3S)$ MC sample, is shown in Fig. \ref{fig_raw_eff}.

\begin{figure}[htpb]
\begin{center}
\epsfig{file={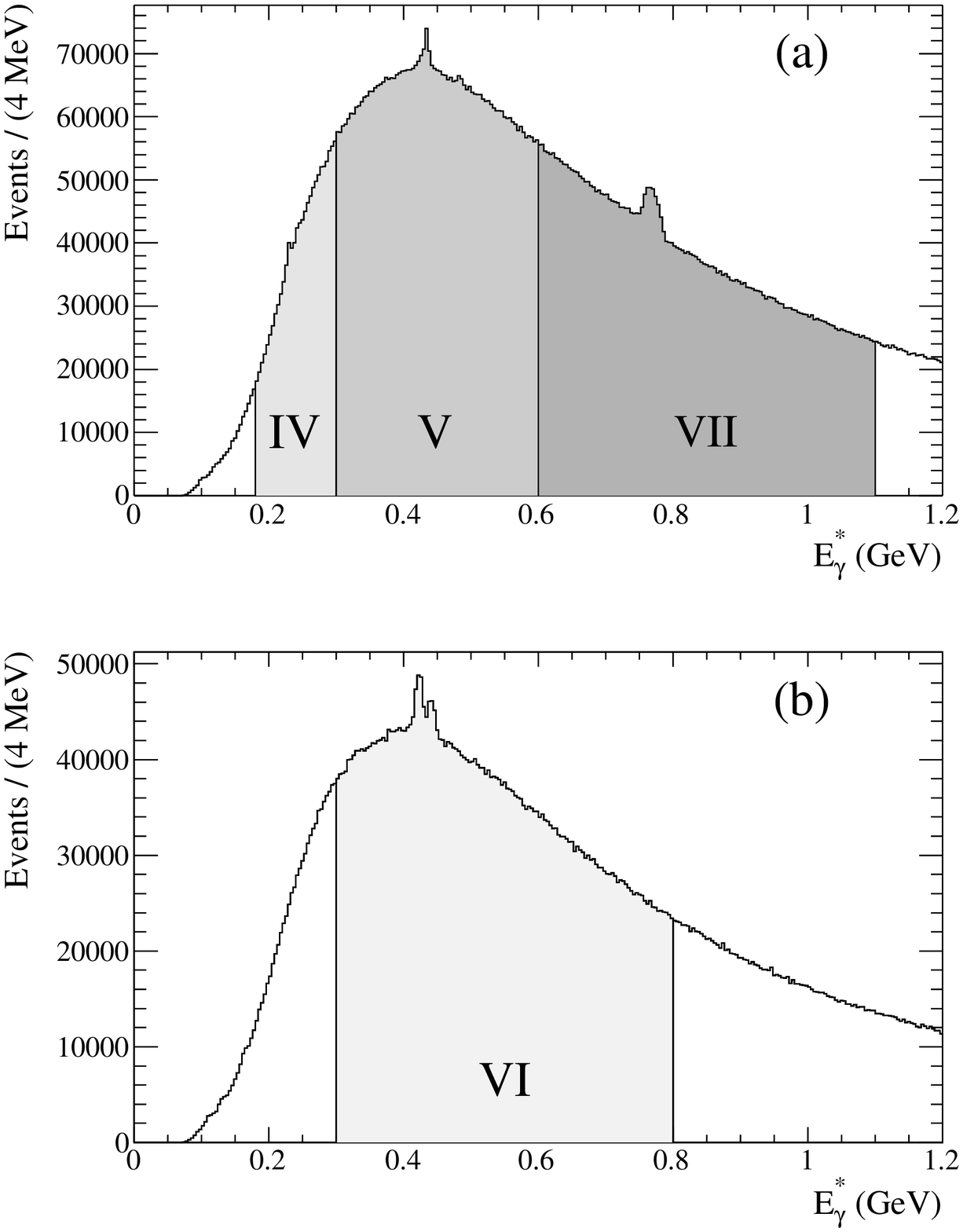},width={0.98\columnwidth},height={1.244\columnwidth}}
\caption{Raw inclusive converted photon energy spectrum from (a) $\Upsilon(3S)$ and (b) $\Upsilon(2S)$ decays. The shaded areas indicate different regions of interest considered in detail in this analysis. The Roman numeral labels indicate the corresponding Section in which each energy region is discussed.}\label{fig_raw}
\end{center}
\end{figure}

\begin{figure}[htpb]
\begin{center}
\epsfig{file={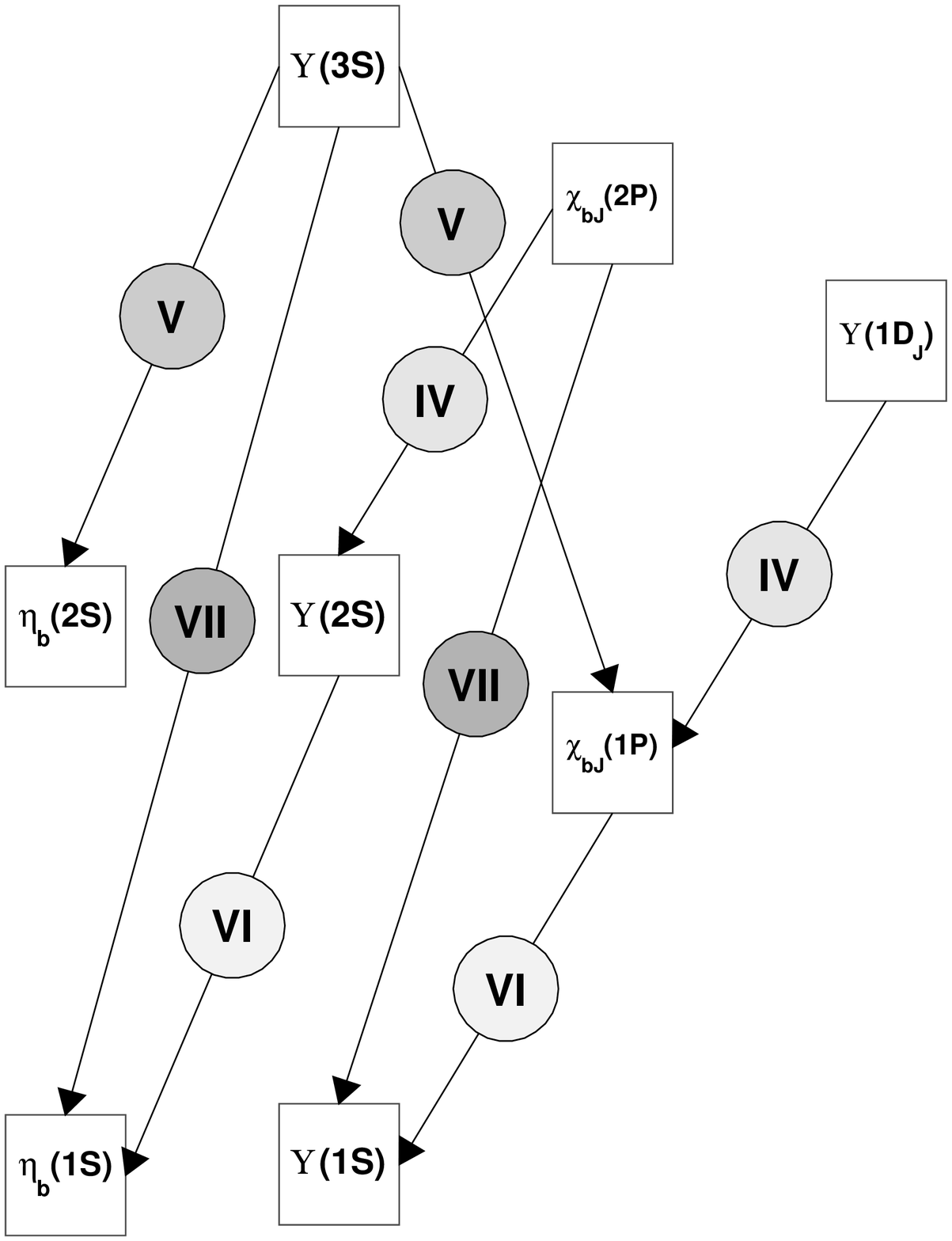},width={0.7\columnwidth},height={0.877\columnwidth}}
\caption{Pictoral representation of energy levels in the bottomonium system and the radiative transitions studied here. The Roman numeral labels indicate the corresponding Section in which the given transition is discussed.}\label{fig_level_diagram}
\end{center}
\end{figure}

Figure \ref{fig_raw} shows the inclusive distributions of the resulting reconstructed converted photon energy.
The data are divided into four energy ranges, as indicated by the shaded regions in Fig. \ref{fig_raw}.
These ranges and the corresponding bottomonium transitions of interest are, in $\Upsilon(3S)$ data:
\begin{itemize}
\item{$180\leq\egamma\leq300$\mev: $\chi_{bJ}(2P)\rightarrow\gamma\Upsilon(2S)$}
\item{$300\leq\egamma\leq600$\mev: $\Upsilon(3S)\rightarrow\gamma\chi_{bJ}(1P)$ and $\Upsilon(3S)\rightarrow\gamma\eta_{b}(2S)$}
\item{$600\leq\egamma\leq1100$\mev: $\chi_{bJ}(2P)\rightarrow\gamma\Upsilon(1S)$ and $\Upsilon(3S)\rightarrow\gamma\eta_{b}(1S)$}
\end{itemize}
and in $\Upsilon(2S)$ data:
\begin{itemize}
\item{$300\leq\egamma\leq800$\mev: $\chi_{bJ}(1P)\rightarrow\gamma\Upsilon(1S)$ and $\Upsilon(2S)\rightarrow\gamma\eta_{b}(1S)$.}
\end{itemize}
Figure \ref{fig_level_diagram} summarizes these energy ranges and the radiative transitions of interest in a pictoral form.
Peaks related to some of these transitions are already clearly visible in Fig. \ref{fig_raw}, where the photon energy in the CM frame of the initial particle for the radiative transition from an initial ($i$) to final ($f$) state is given in terms of their respective masses by
\begin{equation}\label{eqn_egamma}
E_{\gamma} (i\rightarrow f) = \frac{m_{i}^{2}-m_{f}^{2}}{2m_{i}}c^{2}.
\end{equation}
Because we analyse the photon energy in the CM frame of the initial $\Upsilon(mS)$ system (\egamma), the photon spectra from subsequent boosted decays (\emph{e.g.} $\chi_{bJ}(nP)\rightarrow\gamma\Upsilon(1S)$) are affected by Doppler broadening due to the motion of the parent state in the CM frame.

To best enhance the number of signal ($S$) to background ($B$) events, the event selection criteria are chosen by optimizing the figure of merit $\mathcal{F}=\frac{S}{\sqrt{S+B}}$.
This is done separately for each energy region.
The $180\leq\egamma\leq300$\mev energy region in $\Upsilon(3S)$ uses the same criteria as determined for the similarly low energy $300\leq\egamma\leq600$\mev range.
We determine $S$ from MC samples of $\Upsilon(mS)\rightarrow\gamma\etab$ weighted to match the measured branching fractions \cite{pdg}, and $\Upsilon(3S)\rightarrow\gamma\eta_{b}(2S)$ assuming the same branching fraction as for the decay to $\gamma\etab$.
Because the generic decay processes of $\Upsilon(3S)$ are not well-known (ie: a large percentage of the exclusive branching fractions have not been measured), the ``test sample'' data are used to estimate $B$. 
The optimization is performed by varying the selection criteria for the total number of tracks in the event ($nTRK$), the absolute value of the cosine of the angle in the CM frame between the photon momentum and the thrust axis ($|\cos\theta_{T}|$) \cite{thrust}, and a $\pi^{0}$ veto excluding converted photons producing an invariant mass ($m_{\gamma\gamma}$) consistent with $m_{\pi^{0}}$ when paired with any other photon (converted or calorimeter-detected) above a minimum energy ($E_{\gamma 2}$) in the event.
A requirement on the ratio of the second and zeroth Fox-Wolfram moments \cite{fox-wolfram} of each event, $R_{2}$, is also applied.
The reason for using these particular variables (indicated in parentheses) is to preferentially select bottomonium decays to hadronic final states ($nTRK$) and to remove photons from continuum background events ($|\cos\theta_{T}|$ and $R_{2}$) and $\pi^{0}$ decays ($m_{\pi^{0}}$ veto).
Table \ref{tab_cuts} summarizes the values for the optimized selection criteria.

\begin{table}
\begin{center}
\caption {Acceptance criteria for converted photon events.}
\begin{tabular}{cccc}
\hline \hline
Variable & \multicolumn{3}{c}{$E_{\gamma}^{*}$ Range (\mev)}\\
\hline
& $\Upsilon(3S)$ & $\Upsilon(3S)$ & $\Upsilon(2S)$\\
&$[180,600]$&$[600,1100]$&$[300,800]$\\
\hline
$nTRK$ & $\geq8$ & $\geq8$ & $\geq8$\\
$|\cos\theta_{T}|$ & $<0.85$ & $<0.75$ & $<0.85$\\
$|m_{\gamma\gamma}-m_{\pi^{0}}|$ (\mevcc) & $>10$ & $>20$ & $>20$\\
$E_{\gamma 2}$ (\mev) & $>90$ & $>75$ & $>70$\\ 
$R_{2}$ & $<0.98$ & $<0.98$ & $<0.98$\\
\hline \hline
\end{tabular}
\label{tab_cuts}
\end{center}
\end{table}

The efficiency for reconstruction and selection of signal events ($\epsilon$) is determined from MC simulation.
A dedicated $\epem\rightarrow\mu^{+}\mu^{-}\gamma$ sample is used to study our detector model and converted photon efficiency (discussed in Appendix \ref{appendix}), and the correspondence between simulation and data is found to be in very good agreement.
Once the optimal selection criteria have been applied, $\epsilon \lesssim 1.5\%$ for conversions compared to $\sim40\%$ for photons in the EMC for the energy range of interest in this analysis.
Conversely, a large improvement is gained in photon energy resolution, \emph{e.g.} from $\sim25\mev$ in the calorimeter to $4\mev$ or better with converted photons.
Figure \ref{fig_raw} demonstrates both of these features.
The sharply-peaking structures correspond to bottomonium transitions, and are narrow and well-resolved in this analysis.
Unlike in the photon energy spectrum expected from the EMC \cite{babar_etab1, babar_etab2}, the distribution for converted photons drops with energy.
The efficiency decreases (also seen in Fig. \ref{fig_raw_eff}) due to the inability to fully reconstruct the conversion pair as at least one of the individual track momenta approaches the limit of detector sensitivity.
We are unable to contribute useful new information on transitions expected below \egamma=180 (300)\mev for the $\Upsilon(3S)$ ($\Upsilon(2S)$) analysis, which is why those energy ranges are not considered here.

The number of signal events for a given bottomonium transition is extracted from the data by performing a $\chi^{2}$ fit to the \egamma distribution in 1\mev bins.
The functional form and parameterization for each photon signal is determined from MC samples, as described below.
In general, the lineshape is related to the Crystal Ball function \cite{crystal_ball}, \emph{i.e.} a Gaussian function with a power-law tail.
This functional form is used to account for bremsstrahlung losses of the \epem pair.
Comparisons between simulation and data made on $\epem\rightarrow\epem$ events used for the standard luminosity measurement in {\babar\ } demonstrate that the bremsstrahlung tails of these distributions are found to be well-described.
The underlying smooth inclusive photon background is described by a fourth-order polynominal multiplied by an exponential function.
This functional form adequately describes the background in each separate energy range.

\section{\boldmath{$\Upsilon(3S): 180 \leq E_{\gamma}^{*} \leq 300$} M\MakeLowercase{E}V}\label{sec_y3s_low}
\begin{figure*}
\begin{center}
\subfigure{\epsfig{file={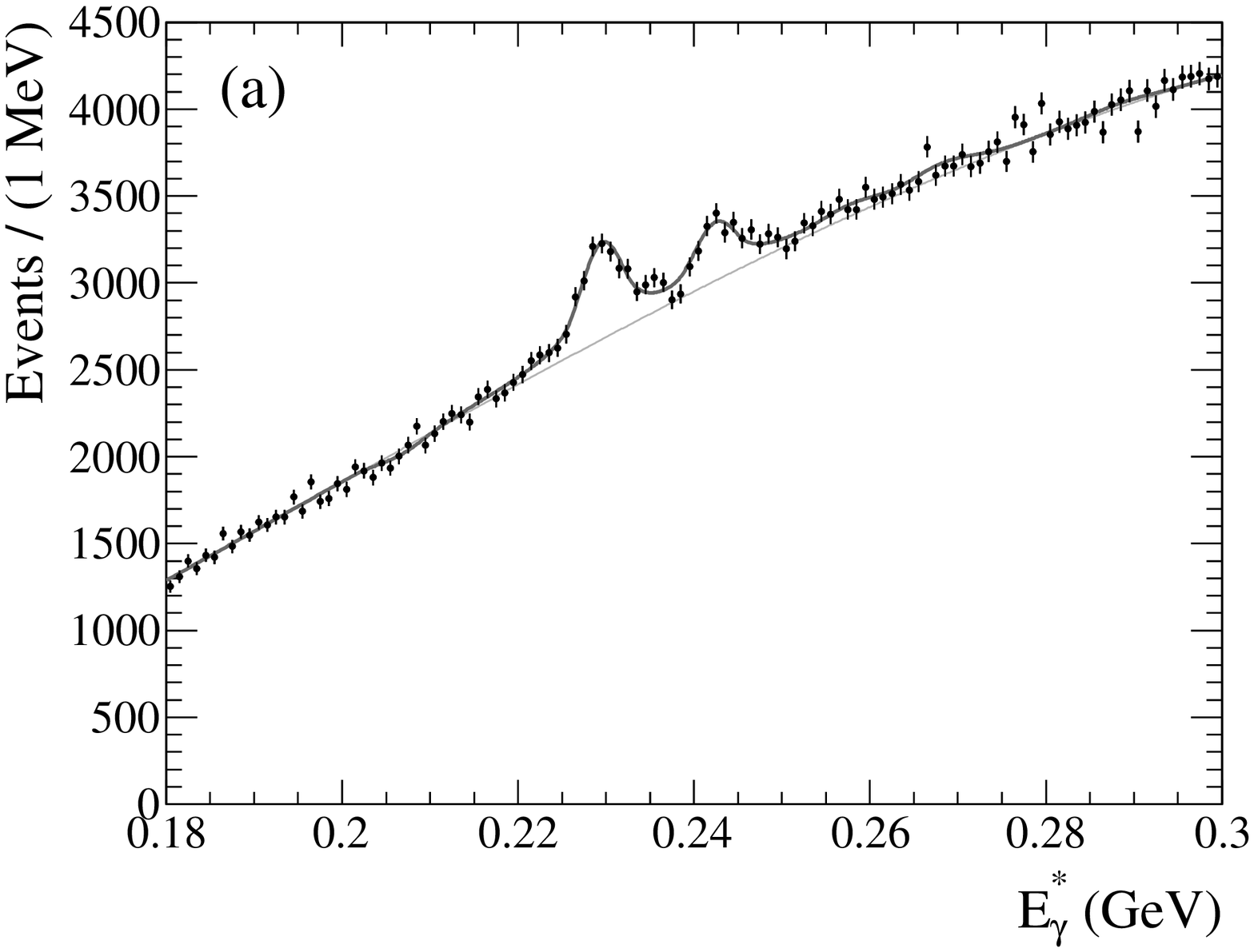},width={1.47\columnwidth},height={1.137\columnwidth}}}
\subfigure{\epsfig{file={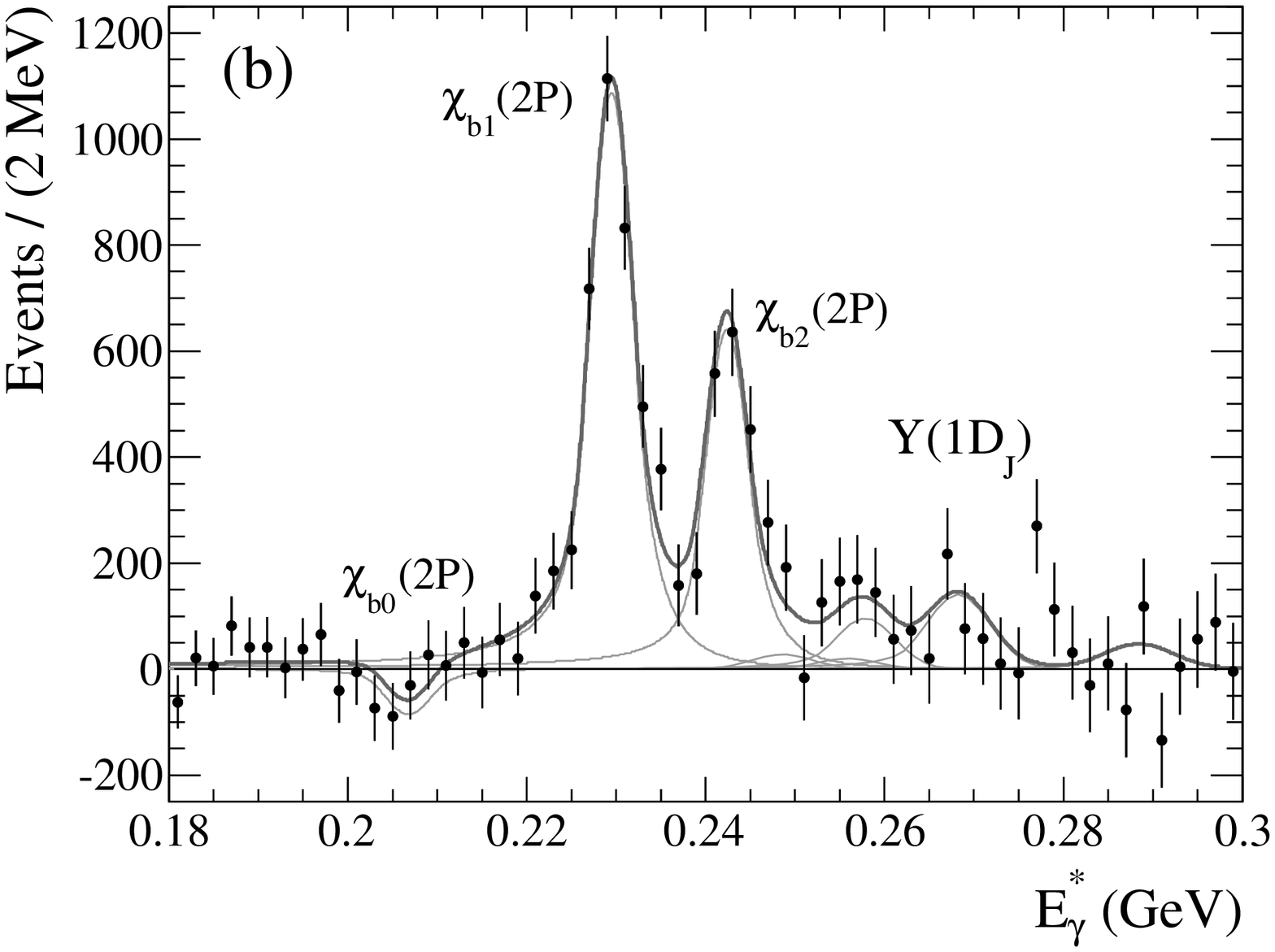},width={1.47\columnwidth},height={1.137\columnwidth}}}
\caption{Fit to the $180 \leq E_{\gamma}^{*} \leq 300$ \mev region of the $\Upsilon(3S)$ data (a) for all of the data, and (b) after subtraction of the fitted background contribution. The $\chi^{2}/ndof$ (where ndof stands for number of degress of freedom) for the fit is 119.1/110. The thin gray lines indicate the individual signal components in the fit, as labelled.} \label{fig_y3s_low}
\end{center}
\end{figure*}

The main purpose of the fit to the $180 \leq E_{\gamma}^{*} \leq 300$\mev region of the $\Upsilon(3S)$ photon energy spectrum, shown in detail in Fig. \ref{fig_y3s_low}, is to measure the $\chi_{bJ}(2P)\rightarrow\gamma\Upsilon(2S)$ transitions.
The only previous measurements of these transitions were made by CUSB \cite{cusb_chib} and CLEO \cite{cleo2} nearly two decades ago.
Those analyses examined the low-energy photon spectrum from exclusive $\Upsilon(3S)\rightarrow\gamma\gamma\Upsilon(2S)(\ell^{+}\ell^{-})$ decays to derive the branching fractions for $\mathcal{B}(\chi_{b1,2}(2P)\rightarrow\gamma\Upsilon(2S))$, and in the case of the CUSB result, to obtain evidence for $\chi_{b0}(2P)\rightarrow\gamma\Upsilon(2S)$.
We present the first fit to \egamma to measure the photon from $\chi_{bJ}(2P)\rightarrow\gamma\Upsilon(2S)$ directly.
Though this analysis is potentially sensitive to all six $\Upsilon(1D_{J})\rightarrow\gamma\chi_{bJ}(1P)$ decays, we treat these decays as a small systematic effect to the $\chi_{bJ}(2P)\rightarrow\gamma\Upsilon(2S)$ measurement.

The $\chi_{bJ}(2P)$ transition lineshapes are parameterized by a Gaussian with power law tails on both the high and low side.
This is best understood as a ``double-sided'' Crystal Ball function with different transition points and exponents for the high and low tails, but with a common Gaussian mean and standard deviation in the central region.
The effects of Doppler broadening, due to the motion of the $\chi_{bJ}(2P)$ in the CM frame, are small ($\sim2\mev$ width) for these transitions.
The $\Upsilon(1D_{J})$-related lineshapes are individually parameterized in terms of a single Crystal Ball function.
Parameterization of these transitions presents a complication because only the mass of the $J=2$ state has been measured reliably \cite{cleo_y1d, babar_y1d}, the value $m_{\Upsilon(1D)}=(10163.7\pm1.4)$ \mevcc being obtained when the experimental results are averaged.
Marginal evidence for the $J=1$ and 3 states was also seen at $\sim10152$\mevcc and $\sim10173$\mevcc, respectively \cite{cleo_y1d, babar_y1d}.
These values are consistent with several theoretical predictions \cite{y1d_godfrey_rosner}, given a shift to bring the theoretical value for $m_{\Upsilon(1D_{2})}$ into agreement with experiment.
We therefore assume the $m_{\Upsilon(1D_{1,3})}$ mass values stated above to compute the expected energy for transitions from those states.
The event yields for these transitions are fixed to the branching fractions expected when $\mathcal{B}(\Upsilon(3S)\rightarrow\gamma\chi_{bJ}(2P))$ \cite{pdg} is combined with the predictions for $\mathcal{B}(\chi_{bJ}(2P)\rightarrow\gamma\gamma\chi_{bJ}(1P))$ via $\Upsilon(1D_{J})$ \cite{y1d_kwong_rosner}.
The efficiencies for the $\Upsilon(1D)$ transition signals range from approximately $0.17$ to $0.30\%$, monotonically rising with \egamma.

\begin{table*}
\begin{center}
\caption {Summary of the analysis of the $180 \leq E_{\gamma}^{*} \leq 300$ \mev region of the $\Upsilon(3S)$ data. The \egamma column lists the transition energy assumed in this analysis. Errors on the yield are statistical only. Regarding the derived $\mathcal{B}(\chi_{bJ}(2P)\rightarrow\gamma\Upsilon(2S))$: the {\babar\ }value is from this paper, while the CUSB and CLEO columns are derivations based on \cite{cusb_chib} and \cite{cleo2} using up-to-date secondary branching fractions from \cite{pdg}. For the {\babar\ }result, the listed uncertainties are statistical, systematic, and from the uncertainties on secondary branching fractions, respectively. For the other results, the total uncertainty (all sources combined in quadrature) is given. Upper limits are given at the $90\%$ confidence level.}
\begin{tabular}{ccccccc}
\hline \hline
Transition & \egamma & Yield & $\epsilon$ & \multicolumn{3}{c}{Derived Branching Fraction $(\%)$}\\
& (\mev) & & ($\%$) & {\babar} & CUSB & CLEO \\
\hline
$\chi_{b0}(2P)\rightarrow\gamma\Upsilon(2S)$ & 205.0 & $-347\pm209$ & 0.105 & $-4.7\pm2.8^{+0.7}_{-0.8}\pm0.5 \:(<2.8)$ & $3.6\pm1.6$ & $<5.2$\\
$\chi_{b1}(2P)\rightarrow\gamma\Upsilon(2S)$ & 229.7 & $4294\pm251$ & 0.152 & $18.9\pm1.1\pm1.2\pm1.8$ & $13.6\pm2.4$ & $21.1\pm4.5$\\
$\chi_{b2}(2P)\rightarrow\gamma\Upsilon(2S)$ & 242.3 & $2462\pm243$ & 0.190 & $8.3\pm0.8\pm0.6\pm1.0$ & $10.9\pm2.2$ & $9.9\pm2.7$\\
\hline \hline
\end{tabular}
\label{tab_y3s_low}
\end{center}
\end{table*}

Figure \ref{fig_y3s_low} shows the measured photon spectrum and results of the fit, before and after subtraction of the inclusive background.
In this fit, the parameters describing the background and any systematic offset in the \egamma scale are free parameters, together with the signal yields for $\chi_{bJ}(2P)\rightarrow\gamma\Upsilon(2S)$ decays.
Table \ref{tab_y3s_low} summarizes the fit results.
Considering both statistical and systematic uncertainties, we find significant $\chi_{b1,2}(2P)\rightarrow\gamma\Upsilon(2S)$ signals ($>12\sigma$ and $>8\sigma$, respectively, where $\sigma$ represents standard deviation), but do not find evidence for $\chi_{b0}(2P)\rightarrow\gamma\Upsilon(2S)$ decay.
The overall energy offset, determined predominantly by the position of the $\chi_{b1,2}(2P)$ transition peaks compared to the nominal \cite{pdg} values, is found to be inconsequential ($-0.3\pm0.2$\mev).

The systematic uncertainties on these measurements (with their approximate sizes given in parentheses below and throughout) include the uncertainty in the fit parameters fixed from MC, uncertainty in the converted photon efficiency, assumptions related to the $\Upsilon(1D_{J})$ contributions, uncertainty on masses used to calculate the expected \egamma values, the $\Upsilon(mS)$ counting uncertainty, effects of the fit mechanics, and the effect of the choice for the background shape.
For each fit component, all of the parameters fixed to MC-determined values are varied individually by $\pm 1\sigma$ of the statistical uncertainty from the MC determination, and the fit repeated.
The maximal variation of the fit result for each component is taken as the systematic uncertainty, and summed in quadrature ($\sim4\%$).
The systematic uncertainty on the converted photon efficiency ($4.7\%$) is estimated using an off-peak control sample and varied selection criteria, as described for all energy regions in Appendix \ref{appendix}.
The fits are repeated with the $\Upsilon(1D_{J})$ masses individually varied by their approximate experimental uncertainties ($\pm1.8$, $\pm1.4$, and $\pm1.5$ \mevcc for $J=1,2$ and 3, respectively) \cite{babar_y1d}, and the fixed yields by $\pm50\%$ of the theoretical values \cite{y1d_kwong_rosner}.
To make a theory-independent determination of the impact due to $\Upsilon(1D_{J})$, the fit is also repeated with four of the $\Upsilon(1D_{J})\rightarrow\gamma\chi_{bJ}(1P)$ yields free to vary (the $\Upsilon(1D_{1})\rightarrow\gamma\chi_{b1}(1P)$ and $\Upsilon(1D_{3})\rightarrow\gamma\chi_{b2}(1P)$ yields are fit as a single component because their \egamma values are nearly identical, and the $\Upsilon(1D_{1})\rightarrow\gamma\chi_{b2}(1P)$ transition is overwhelmed by the main $\chi_{b1,2}(2P)\rightarrow\gamma\Upsilon(2S)$ peaks and remains fixed).
Under this scenario, none of the $\Upsilon(1D_{J})$-related transitions is found to be significant, and the yields are consistent with the theoretical predictions within statistical uncertainty.
The $\chi_{bJ}(2P)\rightarrow\gamma\Upsilon(2S)$ yields are not significantly affected.
The changes in the fit yields for all of these alternative cases are added in quadrature and taken as the systematic uncertainty due to $\Upsilon(1D_{J})$ decays ($\sim2\%$).
It is worth reiterating that the excellent resolution obtained by using converted photons separates the $\Upsilon(1D_{J})$- and $\chi_{bJ}(2P)$-related components in \egamma, which is why the impact of the $\Upsilon(1D_{J})$ states does not dominate the measurement uncertainty.
The fit is repeated with the bottomonium masses (hence, \egamma values) varied according to the PDG uncertainties \cite{pdg}, and the change in the yield added in quadrature ($\sim2\%$).
The number of $\Upsilon(mS)$ mesons and its uncertainty ($1.0\%$) were calculated separately, based on visible cross sections computed from dedicated $\epem\rightarrow\epem(\gamma)$ and $\epem\rightarrow\mu^{+}\mu^{-}(\gamma)$ control samples.
Systematic effects due to the fit mechanics were tested by repeating the fit separately with an expanded \egamma range and a bin width of $0.5\mev$, the difference in results defining a small systematic uncertainty ($1.5\%$).
As a cross-check, the fit was repeated with the $\chi_{b0}(2P)$ component restricted to a physical range.
The effect on the other signal yields was found to be small ($<2\%$).
Finally, the background shape was replaced by a fifth-order polynomial and half of the resulting change in the yield ($<1\%$) taken as the symmetric error due to this assumed parameterization.

We find $\mathcal{B}(\Upsilon(3S)\rightarrow\gamma\chi_{bJ}(2P))\times\mathcal{B}(\chi_{bJ}\rightarrow\gamma\Upsilon(2S))$ = $(-0.3\pm0.2^{+0.5}_{-0.4})\%$, $(2.4\pm0.1\pm0.2)\%$, and $(1.1\pm0.1\pm0.1)\%$ for $J=0$, 1, and 2, respectively.
Using $\mathcal{B}(\Upsilon(3S)\rightarrow\gamma\chi_{bJ}(2P))$ from \cite{pdg}, we derive $\mathcal{B}(\chi_{bJ}(2P)\rightarrow\gamma\Upsilon(2S))$ = $(-4.7\pm2.8^{+0.7}_{-0.8}\pm0.5)\%$, $(18.9\pm1.1\pm1.2\pm1.8)\%$, and $(8.3\pm0.8\pm0.6\pm1.0)\%$, where the errors are statistical, systematic, and from the uncertainty on $\mathcal{B}(\Upsilon(3S)\rightarrow\gamma\chi_{bJ}(2P))$, respectively.
From these values, we calculate a $90\%$ confidence level upper limit of $\mathcal{B}(\chi_{b0}(2P)\rightarrow\gamma\Upsilon(2S))<2.8\%$ \cite{ul}.
Past experimental results \cite{cusb_chib,cleo2} averaged by the PDG \cite{pdg} rely on assumptions for the branching fractions of $\Upsilon(2S)\rightarrow\ell^{+}\ell^{-}$ and $\Upsilon(3S)\rightarrow\gamma\chi_{bJ}(2P)$ and their uncertainties that are no longer valid.
In Table \ref{tab_y3s_low}, we have rescaled these previous results using the current values in order to make a useful comparison.
We find our results to be in good agreement with the previous results, and to be the most precise values to date for the $J=$1 and 2 decays.

\section{\boldmath{$\Upsilon(3S): 300 \leq E_{\gamma}^{*} \leq 600$} M\MakeLowercase{E}V}\label{sec_y3s_medium}
\begin{figure*}
\begin{center}
\subfigure{\epsfig{file={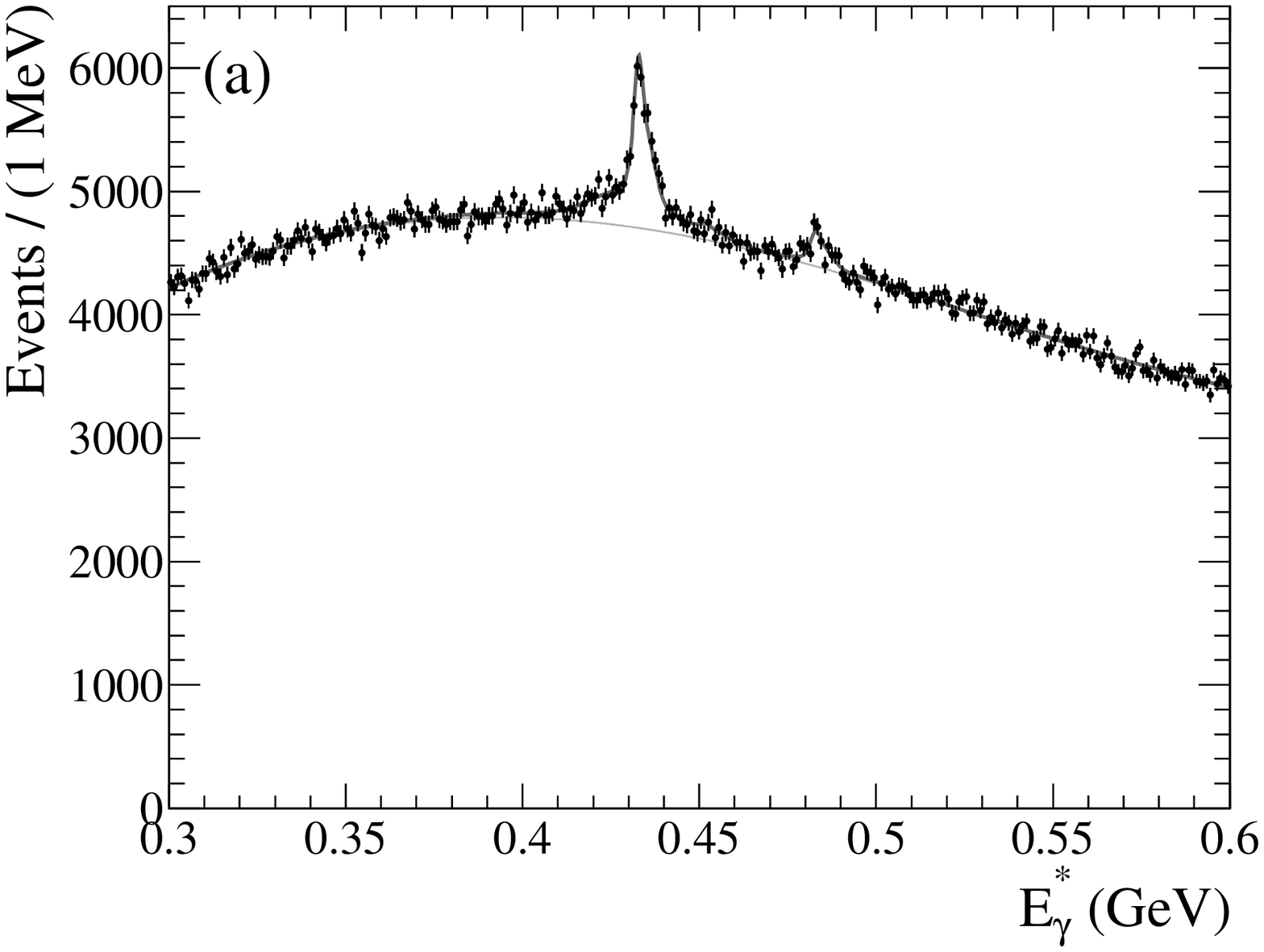},width={1.47\columnwidth},height={1.137\columnwidth}}}
\subfigure{\epsfig{file={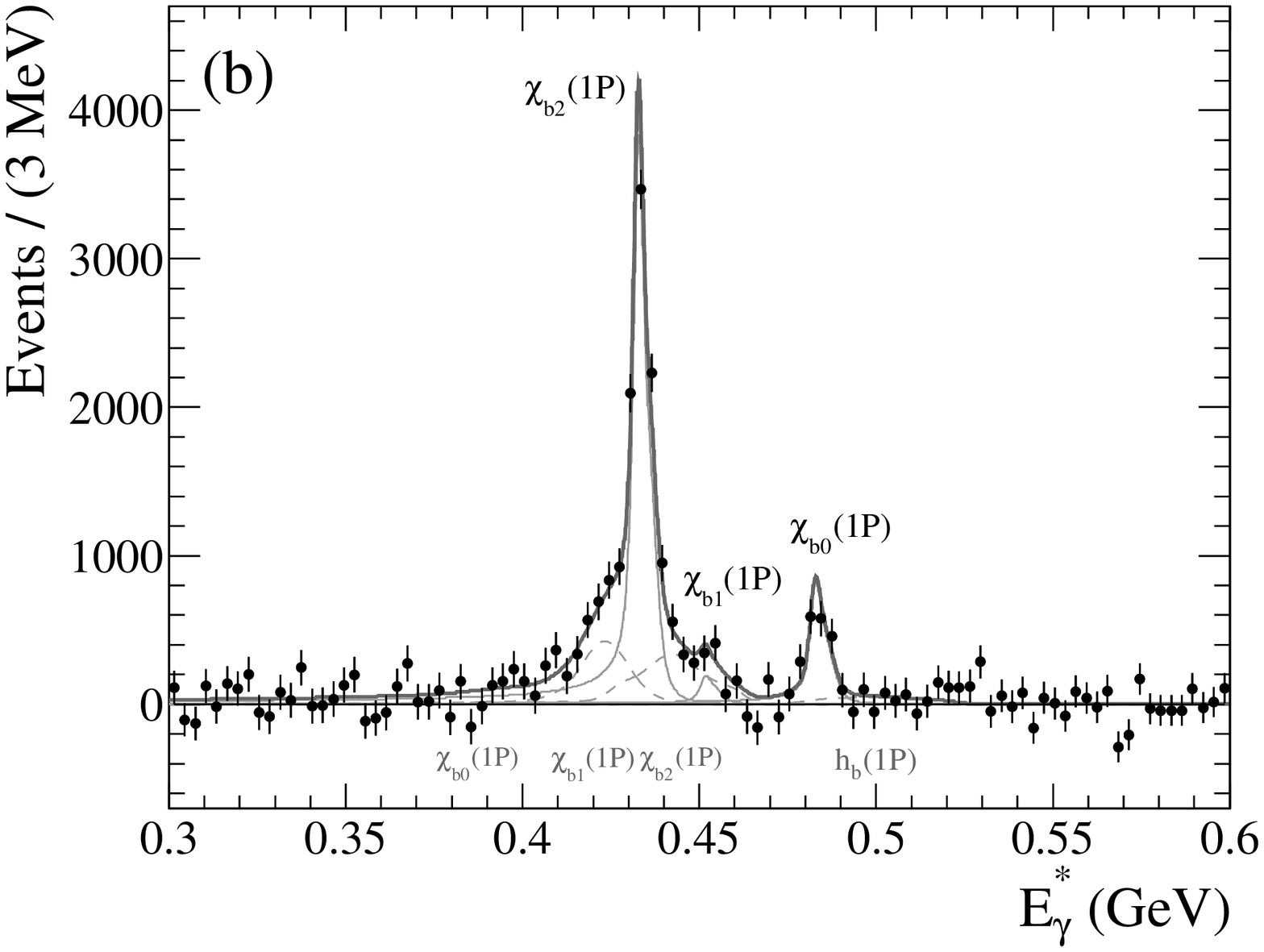},width={1.47\columnwidth},height={1.137\columnwidth}}}
\caption{Fit to the $300 \leq E_{\gamma}^{*} \leq 600$ \mev region of the $\Upsilon(3S)$ data (a) for all of the data, and (b) after subtraction of the fitted background contribution. The $\chi^{2}/ndof$ for the fit is 316/290. The thick solid lines indicate the total fit, the thin solid lines indicate the $\Upsilon(3S)\rightarrow\gamma\chi_{bJ}(1P)$ components, and the dashed lines indicate those from $\chi_{bJ}(1P)\rightarrow\gamma\Upsilon(1S)$ and $h_{b}(1P)\rightarrow\gamma\eta_{b}(1S)$, as labeled.} \label{fig_y3s_med}
\end{center}
\end{figure*}

The $300 \leq E_{\gamma}^{*} \leq 600$\mev range in the inclusive $\Upsilon(3S)$ photon energy spectrum, shown in Fig. \ref{fig_y3s_med}, is complicated by many radiative bottomonium transitions.
A principal feature is the photon lines from the three direct $\Upsilon(3S)\rightarrow\gamma\chi_{bJ}(1P)$ decays.
Photons from the secondary decays, $\chi_{bJ}(1P)\rightarrow\gamma\Upsilon(1S)$, have energies that overlap with these initial transitions.
There are several ways to produce $\chi_{bJ}(1P)$ from $\Upsilon(3S)$, each with unique Doppler broadening and relative rate.
These decays ``feed-down'' to produce many extraneous $\chi_{bJ}(1P)$ mesons that contribute substantially to the background level through subsequent $\chi_{bJ}(1P)\rightarrow\gamma\Upsilon(1S)$ decay.
At the lower edge of this energy range, there are potential contributions from $\Upsilon(3S)\rightarrow\gamma\eta_{b}(2S)$ and $\Upsilon(2S)$ production from initial state radiation (ISR).

The best known of the $\Upsilon(3S)\rightarrow\gamma\chi_{bJ}(1P)$ branching fractions comes from the CLEO experiment, which was able to isolate the $\Upsilon(3S)\rightarrow\gamma\chi_{b0}(1P)$ signal \cite{cleo_inclusive}.
A separate analysis of $\chi_{bJ}(1P)$ decays to multihadronic final states further set upper limits on $\mathcal{B}(\Upsilon(3S)\rightarrow\gamma\chi_{bJ}(1P))$ \cite{cleo_hadrons}.
A recent analysis of $\Upsilon(3S)\rightarrow\gamma\chi_{b1,2}(1P)\rightarrow\gamma\gamma\Upsilon(1S)$ transitions with exclusive $\Upsilon(1S)\rightarrow\ell^{+}\ell^{-}$ decays has resulted in a measurement of $\Upsilon(3S)\rightarrow\gamma\chi_{b1,2}(1P)$ branching fractions \cite{cleo_new}.
Our improved \egamma resolution with the converted photon sample allows us to disentangle the overlapping photon lines to make a direct measurement of these radiative transitions as well.
We also search for a signal for $\Upsilon(3S)\rightarrow\gamma\eta_{b}(2S)$.

The direct $\Upsilon(3S)\rightarrow\gamma\chi_{bJ}(1P)$ lineshapes are parameterized using the double-sided Crystal Ball function described in Sec. \ref{sec_y3s_low} plus an independent Gaussian to account for broadening from non-linearities in the \egamma resolution due to low momentum tracks encountered in this energy range.
The $\Upsilon(3S)\rightarrow\gamma\eta_{b}(2S)$ lineshape is modeled with the convolution of a relativistic Breit-Wigner function (natural lineshape for the $\eta_{b}(2S)$) and a Crystal Ball function (experimental resolution function), where the Breit-Wigner function has been modified by a transformation of variables to \egamma using Eq. (\ref{eqn_egamma}).
The ISR-produced $\Upsilon(2S)$ signal is parameterized with a Crystal Ball function, for which the width is dominated by the spread in the \epem beam energy.

The lineshapes for the decays $\chi_{bJ}(1P)\rightarrow\gamma\Upsilon(1S)$ depend on the initial decays that produced the $\chi_{bJ}(1P)$ states.
We consider six main production pathways:
\begin{itemize}
\item{$\Upsilon(3S)\rightarrow\gamma\chi_{bJ}(1P)$}
\item{$\Upsilon(3S)\rightarrow\gamma\gamma\Upsilon(2S)\rightarrow\gamma\gamma\gamma\chi_{bJ}(1P)$}
\item{$\Upsilon(3S)\rightarrow\gamma\gamma\Upsilon(1D_{J})\rightarrow\gamma\gamma\gamma\chi_{bJ}(1P)$}
\item{$\Upsilon(3S)\rightarrow\pi\pi\Upsilon(2S)\rightarrow\pi\pi\gamma\chi_{bJ}(1P)$}
\item{$\Upsilon(3S)\rightarrow\gamma\chi_{bJ}(2P)\rightarrow\gamma\pi\pi\chi_{bJ}(1P)$}
\item{$\epem\rightarrow\gamma_{ISR}\Upsilon(2S)\rightarrow\gamma_{ISR}\gamma\chi_{bJ}(1P)$.}
\end{itemize}

\begin{table*}
\begin{center}
\caption {Summary of the analysis of the $300 \leq E_{\gamma}^{*} \leq 600$\mev region of the $\Upsilon(3S)$ data. The \egamma column lists the transition energy assumed in this analysis. Errors on the yield are statistical only. For the Derived Branching Fraction, the {\babar\ }values are from this work, and the CLEO results are from \cite{cleo_inclusive, cleo_new}. The upper limit is given at the $90\%$ confidence level.}
\begin{tabular}{cccccc}
\hline \hline
Transition & \egamma & Yield & $\epsilon$ & \multicolumn{2}{c}{Derived Branching Fraction $(\times 10^{-3})$}\\
& (\mev) & & ($\%$) & \babar & CLEO\\
\hline
$\Upsilon(3S)\rightarrow\gamma\chi_{b2}(1P)$ & $433.1$ & $9699\pm318$ & 0.794 & $10.5\pm0.3^{+0.7}_{-0.6}$ & $7.7\pm1.3$\\
$\Upsilon(3S)\rightarrow\gamma\chi_{b1}(1P)$ & $452.2$ & $483\pm315$ & 0.818 & $0.5\pm0.3^{+0.2}_{-0.1} \: (<1.0)$ & $1.6\pm0.5$\\
$\Upsilon(3S)\rightarrow\gamma\chi_{b0}(1P)$ & $483.5$ & $2273\pm307$ & 0.730 & $2.7\pm0.4\pm0.2$ & $3.0\pm1.1$\\ 
\hline \hline
\end{tabular}
\label{tab_y3s_med}
\end{center}
\end{table*}

The feed-down contribution from $\Upsilon(3S)\rightarrow\gamma\chi_{bJ}(1P)$ is determined directly from the fit to the data.
The lineshapes for the subsequent $\chi_{bJ}(1P)\rightarrow\gamma\Upsilon(1S)$ decays are distorted by Doppler-broadening effects.
We parameterize the $\chi_{bJ}(1P)$ transition lineshape with the convolution of a rectangular function and a Crystal Ball function.
Because of the large Doppler width ($\sim20\mev$), the resulting shape is relatively broad and non-peaking.
In the fit, the relative yields of the direct to the secondary transitions are fixed according to the ratios of the expected efficiencies for each mode, and the branching fractions for the $\chi_{bJ}(1P)\rightarrow\gamma\Upsilon(1S)$ decays (to be discussed below).

There are two $3\gamma$ pathways from $\Upsilon(3S)$ to $\chi_{bJ}(1P)$.
Decays via $\Upsilon(2S)$ are fairly well understood, and the precision branching fraction results from Sec. \ref{sec_y3s_low} are used to determine the expected yields and uncertainties.
In contrast, the decays via $\Upsilon(1D_{J})$ have not been measured in detail.
We rely on theoretical predictions \cite{y1d_kwong_rosner}, found to be consistent with an experimental measurement of the $4\gamma$ cascade to $\Upsilon(1S)$ \cite{cleo_y1d}, to estimate the total feed-down component.
We take the uncertainties on $\mathcal{B}(\Upsilon(3S)\rightarrow\gamma\chi_{bJ}(2P))$ \cite{pdg} and introduce a $30\%$ uncertainty on each theoretically calculated branching fraction in the decay chain.
Doppler effects introduce a smooth $\sim 5\mev$ broadening in these (and other) multi-step decay processes, thus the lineshapes for the individual $3\gamma$ pathways are adequately parameterized using a standard Crystal Ball function.

There are two di-pion decay chains leading to $\chi_{bJ}(1P)$: either via $\Upsilon(3S)\rightarrow\pi\pi\Upsilon(2S)$ or $\chi_{bJ}(2P)\rightarrow\pi\pi\chi_{bJ}(1P)$.
The former has been precisely measured by {\babar\ }in a recent analysis of the recoil against $\pi^{+}\pi^{-}$ to search for the $h_{b}(1P)$ state \cite{babar_pipihb}.
We combine the branching fraction from that analysis with the PDG average \cite{pdg} to obtain $\mathcal{B}(\Upsilon(3S)\rightarrow\pi^{+}\pi^{-}\Upsilon(2S)$ = $(2.7\pm0.2)\%$.
For the $\pi^{0}\pi^{0}$ transition, we use the current world average branching fraction value \cite{pdg}.
The relevant MC samples are generated with the experimentally-determined $m_{\pi^{+}\pi^{-}}$ distribution \cite{cleo_dipion_shape}.
Di-pion transitions between $\chi_{bJ}(2P)$ and $\chi_{bJ}(1P)$ for $J=$1 and 2 have been measured experimentally by CLEO \cite{cleo_dipion}.
The above-mentioned {\babar\ }di-pion analysis \cite{babar_pipihb} also measured these quantities, which are averaged with the CLEO results to derive $\mathcal{B}(\chi_{bJ}(2P)\rightarrow\pi^{+}\pi^{-}\chi_{bJ}(1P))$ equal to $(9.1\pm1.0)\times 10^{-3}$ and $(5.0\pm0.6)\times 10^{-3}$ for $J=1$ and 2, respectively.
Decays to the $J=0$ state, with different initial and final $J$ values, and via $\pi^{0}\pi^{0}$ have thus far been below the level of experimental sensitivity.
To calculate the expected feed-down, we assume isospin conservation such that $\Gamma_{\pi^{0}\pi^{0}}=\frac{1}{2}\Gamma_{\pi^{+}\pi^{-}}$, and estimate $\mathcal{B}(\chi_{b0}(2P)\rightarrow\pi\pi\chi_{b0}(1P))$ to be about one-fifth of that of the other $J$ states \cite{theory_dipion}.
We assume a $30\%$ uncertainty on all theoretically-estimated branching fractions.

Radiative decay of ISR-produced $\Upsilon(2S)$ mesons can yield $\chi_{bJ}(1P)$ signals.
The estimated production cross section for $\Upsilon(2S)$ is $(28.6\pm1.4)$ pb \cite{benayoun}, where we have assigned a $5\%$ uncertainty to this theoretical calculation.
We combine this with the $\Upsilon(2S)\rightarrow\gamma\chi_{bJ}(1P)$ branching fraction \cite{pdg} to determine the size of this contribution to the background.
From MC simulation, we conclude that the lineshape may be parameterized with a Crystal Ball function.

Except for feed-down from $\Upsilon(3S)\rightarrow\gamma\chi_{bJ}(1P)$, which is determined from the data, the yields of these components are fixed in the fit.
The branching fractions for the final step of the decay chain, $\mathcal{B}(\chi_{bJ}(1P)\rightarrow\gamma\Upsilon(1S))$, are measured precisely for $J=1$ and 2 in Sec. \ref{sec_y2s_high}.
Our values for these decays are averaged with results from CLEO \cite{cleo_new}.
For decays with $J=0$, the CLEO \cite{cleo_new} Collaboration has recently presented observations.
Since we do not observe this decay in Sec. \ref{sec_y2s_high}, we use the measured branching fraction value from CLEO \cite{cleo_new}.

In the fit, we include two components related to $h_{b}(1P)\rightarrow\gamma\etab$ decays.
The $h_{b}(1P)$ decay is assumed to decay with a large branching fraction via $h_{b}(1P)\rightarrow\gamma\etab$ \cite{hb_godfrey_rosner}.
The two relevant $h_{b}(1P)$ production mechanisms are $\Upsilon(3S)\rightarrow\pi^{+}\pi^{-}h_{b}(1P)$ and $\Upsilon(3S)\rightarrow\pi^{0} h_{b}(1P)$.
{\babar\ }has studied both of these modes, finding $\mathcal{B}(\Upsilon(3S)\rightarrow\pi^{+}\pi^{-}h_{b}(1P))$ $<2.5\times 10^{-4}$ \cite{babar_pipihb} and $\mathcal{B}(\Upsilon(3S)\rightarrow\pi^{0} h_{b}(1P))\times\mathcal{B}(h_{b}(1P)\rightarrow\gamma\etab)$ = $(4.7\pm1.5\pm0.6)\times 10^{-4}$ \cite{babar_pi0hb}.
Due to the effects of Doppler broadening, we parameterize the decay via $\pi^{0}$ using the Doppler-broadened Crystal Ball function as described for $\chi_{bJ}(1P)\rightarrow\gamma\Upsilon(1S)$ transitions from $\Upsilon(3S)\rightarrow\gamma\chi_{bJ}(1P)$, and via $\pi^{+}\pi^{-}$ using a standard Crystal Ball function.
The yields for these components are fixed in the fit, and are nearly negligible.

In the fit, all of the lineshape parameters are fixed to the MC-determined values except for the yield of the $\Upsilon(3S)\rightarrow\gamma\chi_{bJ}(1P)$ (and its related $\chi_{bJ}(1P)\rightarrow\gamma\Upsilon(1S)$ components), an overall \egamma scale offset, and the background lineshape parameters.
The feed-down yields are fixed using the branching fractions as described above.
Repeated trials of the signal extraction on simulated datasets determine that, given the low efficiency and expected number of events, and high level of background, obtaining a reliable yield for $\eta_{b}(2S)$ and ISR-produced $\Upsilon(2S)$ is not possible.
These components are therefore not included in the fit.
The measured photon energy spectrum and the fitted yields are presented in Fig. \ref{fig_y3s_med}, before and after the subtraction of the inclusive background.
There is a clear separation of the $\Upsilon(3S)\rightarrow\gamma\chi_{bJ}(1P)$ transitions, enabling us to observe the transitions to $J=0,2$, and find only a very small indication for $J=1$.
Table \ref{tab_y3s_med} summarizes the fit results.

We consider systematic uncertainties due to the choice of background shape ($1-2\%$), fit range, and binning ($1.5\%$), the effect of fixing parameters to the MC-determined values ($3-6\%$), uncertainty in the photon conversion efficiency ($3.6\%$), uncertainty in the $\Upsilon(mS)$ counting ($1.0\%$), uncertainty in the bottomonium masses ($1-3\%$), and the impact of fixed feed-down yields ($2\%$).
The values in parentheses are representative of the $\Upsilon(3S)\rightarrow\gamma\chi_{b0,2}(1P)$ decays; for the $\chi_{b1}$-related results, the effects of the feed-down lineshapes and the yields and the background shape dominate (about $20\%$ each) due the marginal signal size.
The evaluation of these uncertainties is done as described in Sec. \ref{sec_y3s_low}, with the exception of the feed-down-related uncertainty that is unique to this energy region.
To assess the uncertainty related to the assumed branching fractions, we repeat the analysis many times with the value of each input branching fraction varied randomly within its total uncertainty.
We adopt the standard deviation of the change in the results as a systematic error.
As a cross-check, we repeat the fit with the yields of the $\Upsilon(1D)$-related feed-down components allowed to vary as a free parameter.
We find only a small change ($<2\%$) in the overall branching fraction results, and consider this to be sufficiently accounted for by the systematic uncertainty determined from our procedure of varying the branching fractions.
Including ISR and $\eta_{b}(2S)$ components in the fit produces an effect of less than $\sim1\%$, due to their slight impact on determining the overall background shape.

We measure $\mathcal{B}(\Upsilon(3S)\rightarrow\gamma\chi_{bJ}(1P))$ = $(2.7\pm0.4\pm0.2)\times 10^{-3}$, $(0.5\pm0.3^{+0.2}_{-0.1})\times 10^{-3}$, and $(10.5\pm0.3^{+0.7}_{-0.6})\times 10^{-3}$ for $J=0,1$ and 2, respectively.
We observe evidence for the $\Upsilon(3S)\rightarrow\gamma\chi_{b0,2}(1P)$ transitions, with total significances greater than $6.8\sigma$ and $16\sigma$, respectively.
We do not find evidence for the suppressed $\Upsilon(3S)\rightarrow\gamma\chi_{b1}(1P)$ decay, and set the $90\%$ confidence level upper limit of $\mathcal{B}(\Upsilon(3S)\rightarrow\gamma\chi_{b1}(1P))<1.0\times 10^{-3}$.
These results are consistent with previous limits \cite{cleo_hadrons}, and improve upon the only measured value for the $J=0$ transition \cite{cleo_inclusive}.
Our measurements of the $\Upsilon(3S)\rightarrow\gamma\chi_{b1,2}(1P)$ branching fractions both differ from the recent CLEO observations \cite{cleo_new} by nearly $2\sigma$.
Forcing the $\chi_{b1,2}(1P)$ yields in our fit to match the CLEO results gives a poor $\chi^{2}/ndof$ of 399/293. 
However, using the $\mathcal{B}(\chi_{b1,2}(1P)\rightarrow\gamma\Upsilon(1S))$ results from Sec. \ref{sec_y2s_high} to derive a total $\Upsilon(3S)\rightarrow\gamma\gamma\Upsilon(1S)$ branching fraction via $\chi_{b1,2}(1P)$ (comparable to ``$J=$1 and 2'' \cite{cleo_new}), we find the results of the two experiments to be in close agreement.

Adopting these results, we search for the $\Upsilon(3S)\rightarrow\gamma\eta_{b}(2S)$ transition in the range $335\leq\egamma\leq375$\mev and find no evidence.
Taking into account the dominant statistical uncertainty, we derive an upper limit of $\mathcal{B}(\Upsilon(3S)\rightarrow\gamma\eta_{b}(2S))<1.9\times 10^{-3}$ at the $90\%$ confidence level.
This limit is a factor of two larger than the limit set by CLEO \cite{cleo_inclusive}.

\section{\boldmath{$\Upsilon(2S): 300 \leq E_{\gamma}^{*} \leq 800$} M\MakeLowercase{E}V}\label{sec_y2s_high}
\begin{figure*}
\begin{center}
\subfigure{\epsfig{file={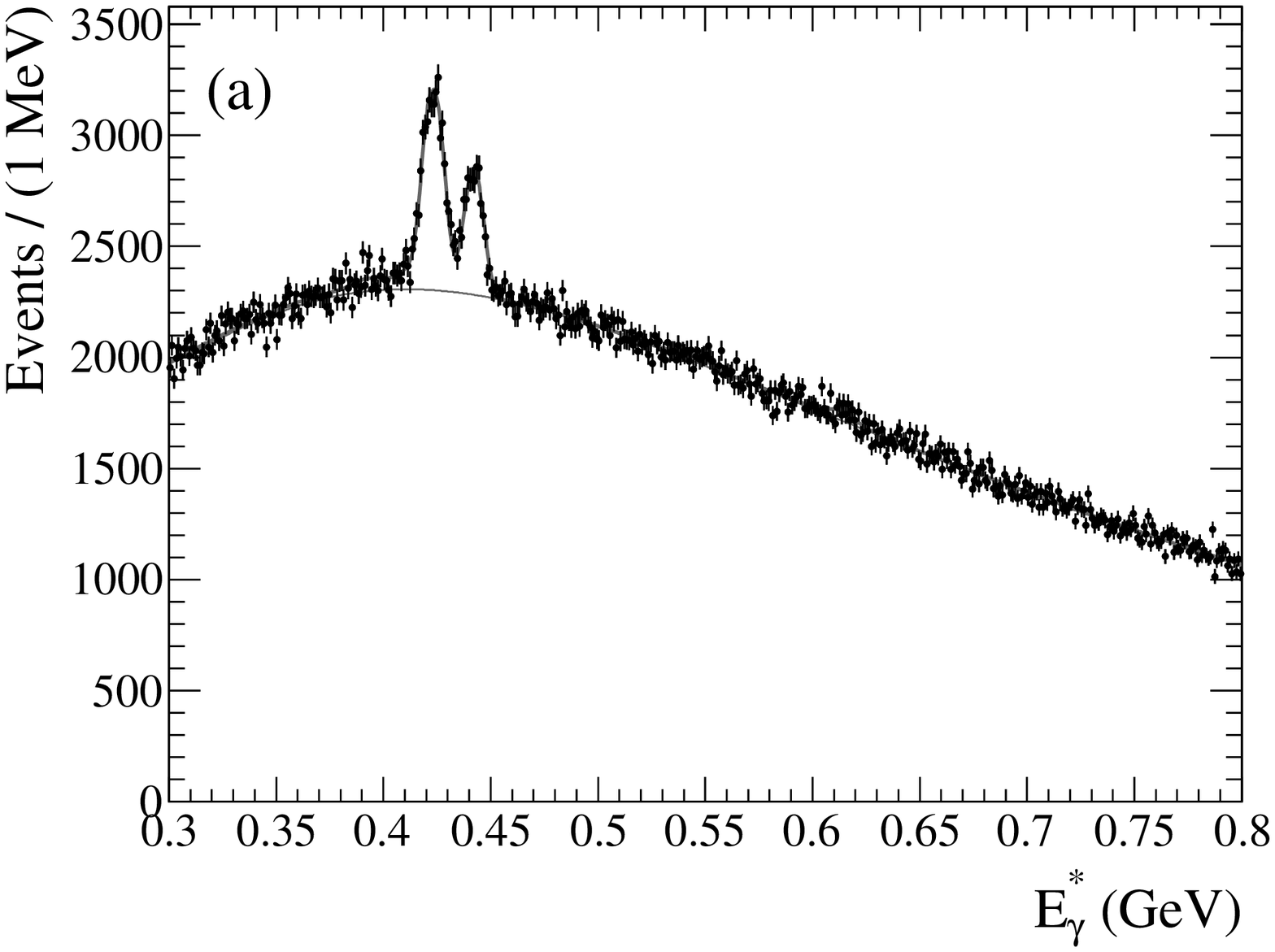},width={1.47\columnwidth},height={1.137\columnwidth}}}
\subfigure{\epsfig{file={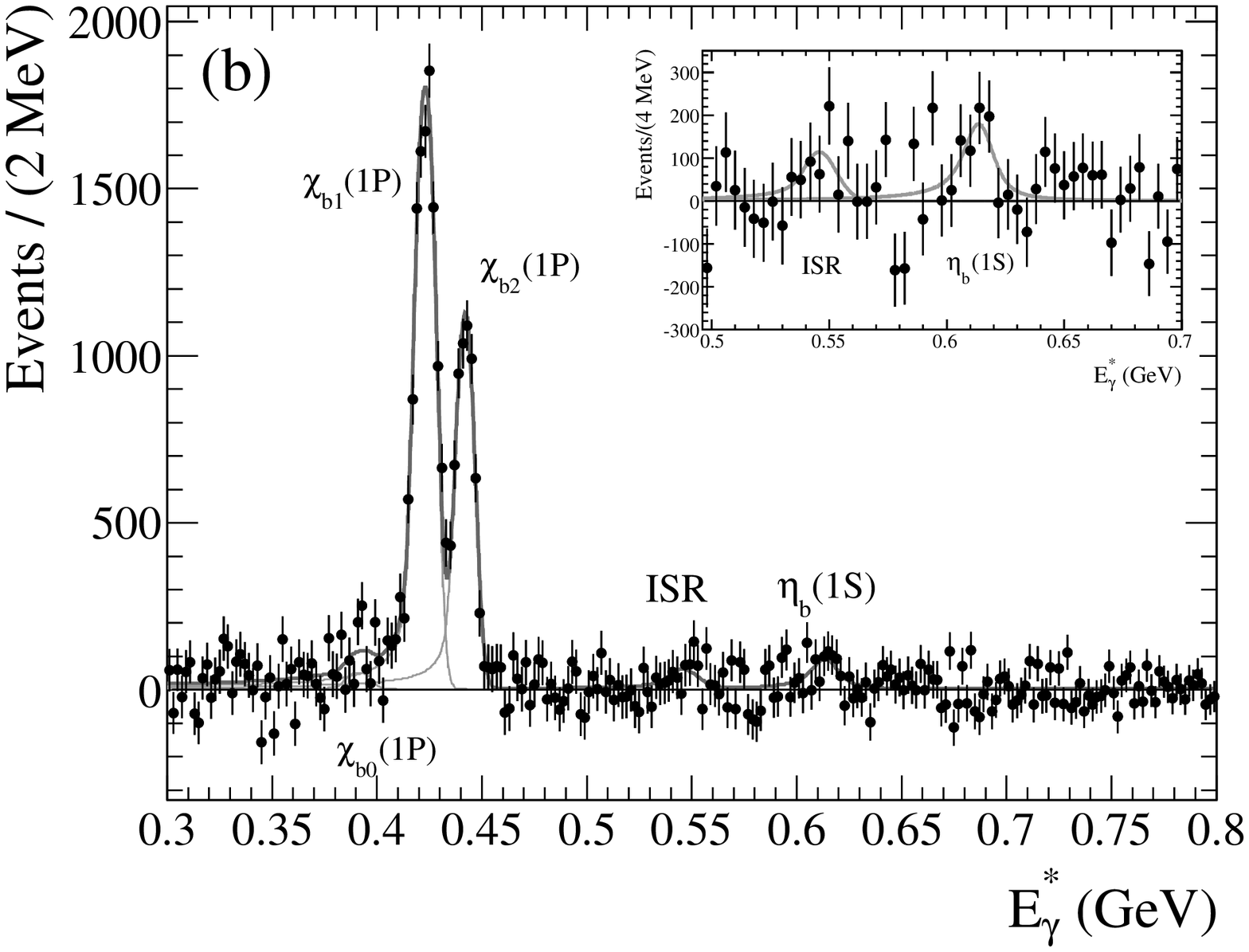},width={1.47\columnwidth},height={1.137\columnwidth}}}
\caption{Fit to the $300 \leq E_{\gamma}^{*} \leq 800$ \mev region of the $\Upsilon(2S)$ data (a) for all of the data, and (b) after subtraction of the fitted background contribution, where the inset focuses on the $\Upsilon(2S)\rightarrow\gamma\etab$ region of the fit. The thin lines indicate the individual fit components. For this fit, $\chi^{2}/ndof=511.0/487$.} \label{fig_y2s_high}
\end{center}
\end{figure*}

We study five possible signals in the $300 \leq E_{\gamma}^{*} \leq 800$ \mev range in $\Upsilon(2S)$ data: three $\chi_{bJ}(1P)\rightarrow\gamma\Upsilon(1S)$ transitions, ISR $\Upsilon(1S)$ production, and $\Upsilon(2S)\rightarrow\gamma\etab$.
This energy region, shown in Fig. \ref{fig_y2s_high}, has been analysed using calorimeter-detected photons by both {\babar\ }\cite{babar_etab2} and CLEO \cite{cleo_etab}, the former finding evidence to confirm the \etab.
The improvement in resolution from the converted photon sample could allow a precise measurement of the \etab mass.
However, because \egamma for the $\Upsilon(2S)\rightarrow\gamma\etab$ transition is $\approx613$\mev (compared to $\approx920$\mev in the $\Upsilon(3S)$ data), its measurement is more difficult due to a lower detection efficiency and larger inclusive photon background.
Studying this energy range is nonetheless useful, since the branching fractions for $\chi_{bJ}(1P)\rightarrow\gamma\Upsilon(1S)$ have had large uncertainties \cite{cb2,cusb1,cusb2} until very recently \cite{cleo_new}, and the values are necessary inputs to the analysis described in Sec. \ref{sec_y3s_medium}.
The $J=0$ decay has also only been recently observed \cite{cleo_new, belle_chib0}.
These external measurements were unavailable when this analysis was initiated.

We parameterize the $\chi_{bJ}(1P)$ transition lineshape with a Doppler-broadened Crystal Ball function, as described in Sec. \ref{sec_y3s_medium}.
The ISR and $\Upsilon(2S)\rightarrow\gamma\etab$ lineshapes are modeled with a Crystal Ball function, and relativistic Breit-Wigner function convolved with a Crystal Ball function, respectively.
The lineshape parameters are determined from MC samples.
Several different natural widths are tested for the \etab, and because the Crystal Ball parameter values (related to \egamma resolution) are found to be independent of the width, the values averaged over all samples are used.
In the fit to the data, all of the parameters are fixed to these MC-determined values, except for the yields for the $\chi_{bJ}(1P)$, ISR, and \etab signals, the mass of the \etab, the inclusive background shape parameters, and an overall \egamma scale offset.
The width of \etab is fixed to 10 \mev.

Figure \ref{fig_y2s_high} shows the converted photon energy spectrum before and after the subtraction of the inclusive background, with an inset focusing on the region of the expected $\Upsilon(2S)\rightarrow\gamma\etab$ transition.
The \egamma resolution provides clear separation of the $\chi_{b1,2}(1P)$-related peaks, allowing for the first direct measurement of these transitions in an inclusive sample.
The results of the fit are summarized in Table \ref{tab_y2s_high}.
We find no evidence for $\chi_{b0}(1P)\rightarrow\gamma\Upsilon(1S)$ decay.
The $\Upsilon(1S)$ yield from ISR production is consistent, within large uncertainties, with the result scaled from the previous {\babar\ }measurement \cite{babar_etab2}.
As expected from signal extraction studies on simulated datasets, the search for a signal in the $\Upsilon(2S)\rightarrow\gamma\etab$ energy region does not find a reliable result.
Estimating the statistical significance from the change in $\chi^{2}$ of the fit with and without this component results in the equivalent of a less than $2.5\sigma$ effect.
The \egamma scale offset in this energy range is $-0.9^{+0.5}_{-0.4}$\mev.

\begin{table*}
\begin{center}
\caption {Summary of the analysis of the $300 \leq E_{\gamma}^{*} \leq 800$ \mev region of the $\Upsilon(2S)$ data. The \egamma column lists the transition energy assumed in this analysis, or in the case of $\Upsilon(2S)\rightarrow\gamma\etab$, the most significant $(\sim1.7\sigma)$ feature in the relevant \egamma region. Errors on the yield are statistical only. Regarding the derived branching fractions, the {\babar\ }value is from this paper, while the Crystal Ball (CB) and CUSB columns are derivations based on Ref. \cite{cb2} and \cite{cusb1} using up-to-date secondary branching fractions from the PDG \cite{pdg}; the CLEO results are from \cite{cleo_new}. For the {\babar\ }result, the listed uncertainties are statistical, systematic, and from the uncertainties on secondary branching fractions, respectively. Upper limits are given at the $90\%$ confidence level. Dashes indicate that no value has been reported in the relevant reference.}
\begin{tabular}{cccccccc}
\hline \hline
Transition & \egamma & Yield & $\epsilon$ & \multicolumn{4}{c}{Derived Branching Fraction $(\%)$}\\
& (\mev) & & ($\%$) & {\babar} & CB & CUSB & CLEO \\
\hline
$\chi_{b0}(1P)\rightarrow\gamma\Upsilon(1S)$ & $391.5$ & $391\pm267$ & 0.496 & $2.2\pm1.5^{+1.0}_{-0.7}\pm0.2 \: (<4.6)$ & $<5$ & $<12$ & $1.7\pm0.4$\\
$\chi_{b1}(1P)\rightarrow\gamma\Upsilon(1S)$ & $423.0$ & $12604\pm285$ & 0.548 & $34.9\pm0.8\pm2.2\pm2.0$ & $34\pm7$ & $40\pm10$ & $33.0\pm2.6$\\
$\chi_{b2}(1P)\rightarrow\gamma\Upsilon(1S)$ & $442.0$ & $7665^{+270}_{-272}$ & 0.576 & $19.5\pm0.7^{+1.3}_{-1.5}\pm1.0$ & $25\pm6$ & $19\pm8$ & $18.5\pm1.4$\\
$\Upsilon(2S)\rightarrow\gamma\etab$ & $613.7^{+3.0+0.7}_{-2.6-1.1}$ & $1109\pm348$ & $1.050$ & $0.11\pm0.04^{+0.07}_{-0.05} \: (<0.21)$ & - & - & -\\
\hline \hline
\end{tabular}
\label{tab_y2s_high}
\end{center}
\end{table*}

The systematic uncertainties on these measurements are related to the choice of background shape, the fit mechanics, the effect of fixing parameters to the MC-determined values, uncertainty in the photon conversion efficiency, uncertainty in the $\Upsilon(mS)$ counting, uncertainties in the bottomonium masses, and assumptions on the \etab width.
The methodology for the evaluation of these uncertainties has been described for the most part in Sec. \ref{sec_y3s_low}.
The systematic uncertainty related to the \etab width is estimated by finding the maximal change in yield when the fit is repeated using a range of widths between $2.5-15$\mev, values consistent with a wide range of theoretical predictions.
While varying the assumed \etab width affects the event yield, it is found to have a negligible impact on the significance of the signal.
For the $\chi_{b1,2}(1P)\rightarrow\gamma\Upsilon(1S)$ transitions, the largest sources of uncertainty are related to the fixed lineshape parameters ($3-4\%$), uncertainty in the bottomonium masses ($\sim4\%$ for $\chi_{b2}(2P)$, and dominant for the \egamma scale uncertainty) and the conversion efficiency ($5.2\%$).
Each of the remaining sources contributes less than $2\%$.
For the \etab signal, systematic uncertainties dominate the result.
The largest effects are due to varying the background shape ($\sim31\%$), the bottomonium masses ($\sim25\%$), the MC-determined parameters ($\sim22\%$), and the \etab width ($\sim16\%$).

We measure $\mathcal{B}(\Upsilon(2S)\rightarrow\gamma\chi_{bJ}(1P))\times\mathcal{B}(\chi_{bJ}(1P)\rightarrow\gamma\Upsilon(1S))$ = $(8.3\pm5.6^{+3.7}_{-2.6})\times 10^{-4}$, $(24.1\pm0.6\pm1.5)\times 10^{-3}$, and $(13.9\pm0.5^{+0.9}_{-1.1})\times 10^{-3}$, for $J=$ 0, 1 and 2, respectively.
Using $\mathcal{B}(\Upsilon(2S)\rightarrow\gamma\chi_{bJ}(1P))$ from the PDG \cite{pdg}, we derive $\mathcal{B}(\chi_{bJ}(1P)\rightarrow\gamma\Upsilon(1S))$ = $(2.2\pm1.5^{+1.0}_{-0.7}\pm0.2)\%$, $(34.9\pm0.8\pm2.2\pm2.0)\%$, and $(19.5\pm0.7^{+1.3}_{-1.5}\pm1.0)\%$, where the uncertainties are statistical, systematic, and from the uncertainty on $\mathcal{B}(\Upsilon(2S)\rightarrow\gamma\chi_{bJ}(1P))$, respectively.
We calculate a $90\%$ confidence level upper limit of $\mathcal{B}(\chi_{b0}(1P)\rightarrow\gamma\Upsilon(1S))<4.6\%$.
As previously, we rescale the existing results \cite{cb2,cusb1} using the most up-to-date secondary branching fraction values \cite{pdg} to obtain the results quoted in Table \ref{tab_y2s_high}.
Our $\chi_{bJ}(1P)$ transition results agree with the previous measurements, but represent a two- to three-fold reduction in the total uncertainty.
We find reasonable agreement with, and a comparable precision to, the recent measurements from CLEO \cite{cleo_new}. 
When the yield-related systematic uncertainties on the measurement of the \etab candidate are taken into account (excluding those due to the \etab width), the result is further reduced in significance to an equivalent of $\sim 1.7\sigma$.
We find no evidence for an \etab signal in this analysis of the $\Upsilon(2S)$ dataset, and set a corresponding limit of $\mathcal{B}(\Upsilon(2S)\rightarrow\gamma\etab)<0.21\%$.

\section{\boldmath{$\Upsilon(3S): 600 \leq E_{\gamma}^{*} \leq 1100$} M\MakeLowercase{E}V}\label{sec_y3s_high}
\begin{figure*}
\begin{center}
\subfigure{\epsfig{file={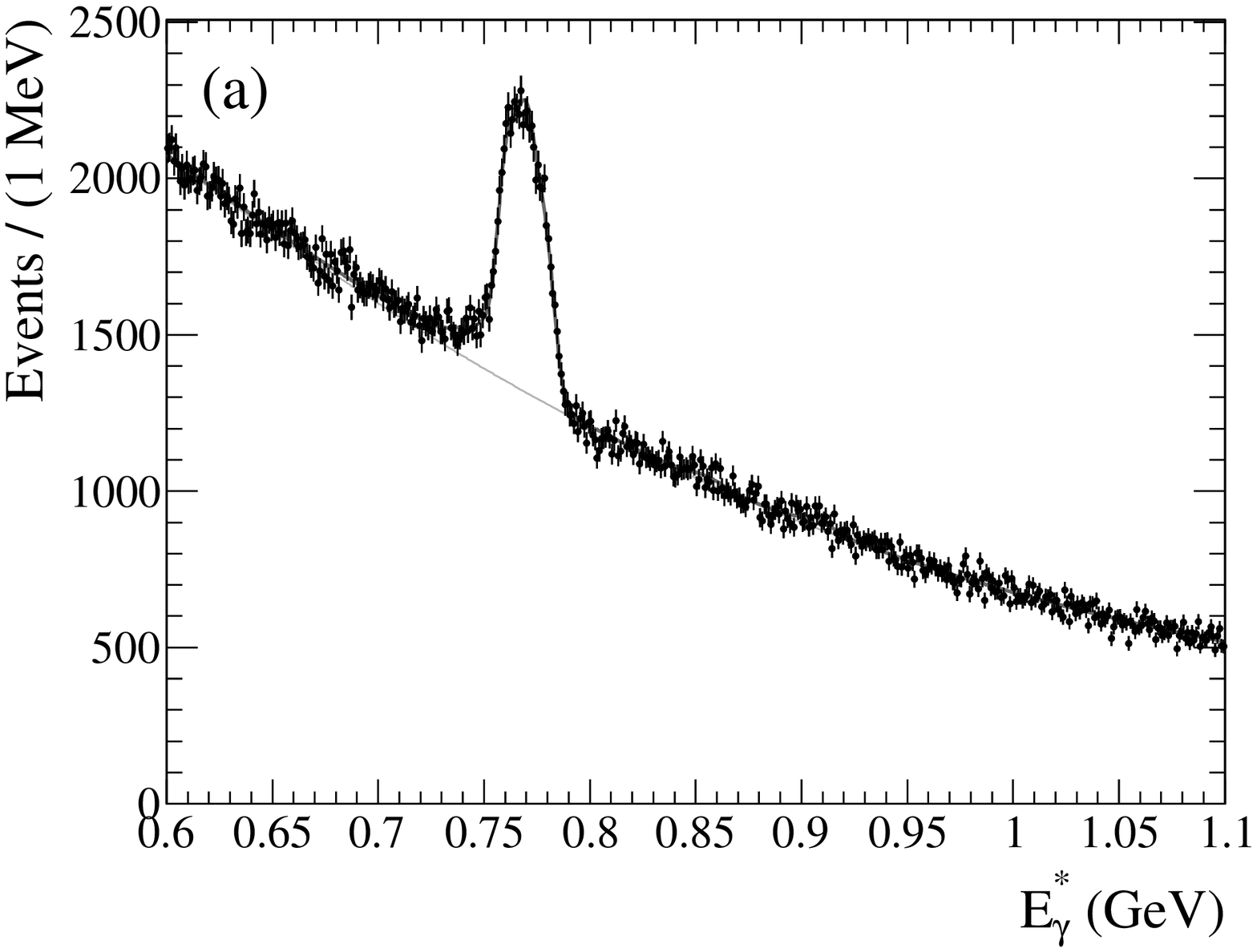},width={1.47\columnwidth},height={1.137\columnwidth}}}
\subfigure{\epsfig{file={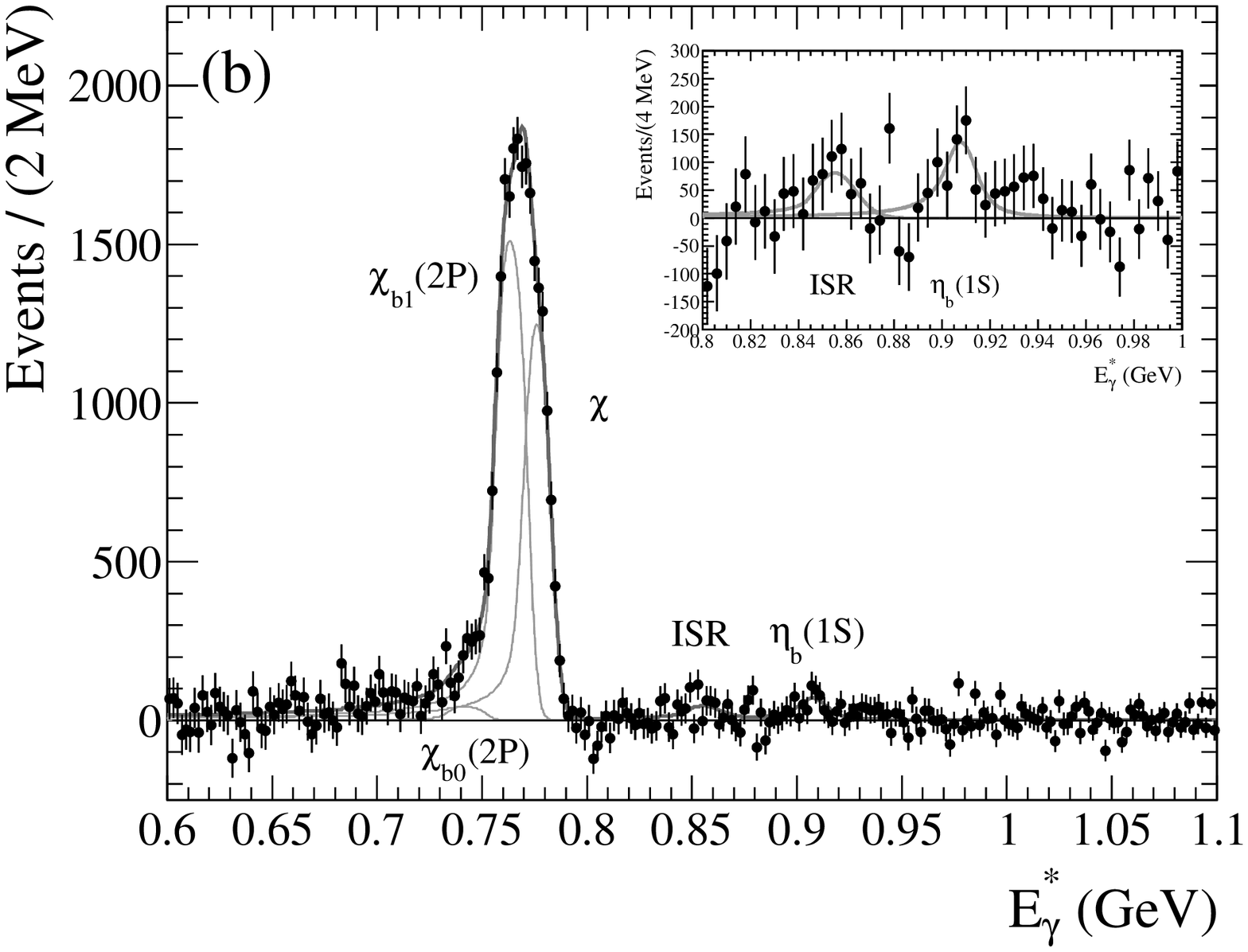},width={1.47\columnwidth},height={1.137\columnwidth}}}
\caption{Fit to the $600 \leq E_{\gamma}^{*} \leq 1100$ \mev region of the $\Upsilon(3S)$ data (a) for all of the data, and (b) after subtraction of the fitted background contribution, where the inset focuses on the $\Upsilon(3S)\rightarrow\gamma\etab$ region of the fit. The thin lines indicate the individual fit components. For this fit, $\chi^{2}/ndof=442.9/487$.} \label{fig_y3s_high}
\end{center}
\end{figure*}

The analysis of the $600 \leq E_{\gamma}^{*} \leq 1100$ \mev region for the $\Upsilon(3S)$, shown in Fig. \ref{fig_y3s_high}, is very similar to that in Sec. \ref{sec_y2s_high} of the $300 \leq E_{\gamma}^{*} \leq 800$ \mev region for the $\Upsilon(2S)$.
Again, we study potential signals from three $\chi_{bJ}(2P)\rightarrow\gamma\Upsilon(1S)$ transitions, $\Upsilon(1S)$ production from ISR, and $\Upsilon(3S)\rightarrow\gamma\etab$.
In this case, the calorimeter-based analysis of the same region produced the discovery of the \etab \cite{babar_etab1}.
The higher \egamma value for $\Upsilon(3S)\rightarrow\gamma\etab$ offers the advantages of both an increased efficiency and lower background level compared to the analagous analysis in $\Upsilon(2S)$ data, and therefore a better sensitivity for the observation of $\eta_{b}(1S)$.
There is also the possibility of updating the measurements of $\chi_{bJ}(2P)\rightarrow\gamma\Upsilon(1S)$ transitions, including confirmation of the decay of the $J=0$ state \cite{cusb_chib, cleo2}.

We parameterize the signal lineshape in the same manner as described in Sec. \ref{sec_y2s_high}, with Doppler-broadened Crystal Ball functions for the $\chi_{bJ}(2P)$ transitions, a Crystal Ball function for ISR production of the $\Upsilon(1S)$, and the relativistic Breit-Wigner Crystal Ball convolution for the $\etab$ signal.
As before, all of the lineshape parameters are fixed to their MC-determined values, with the yields for the $\chi_{bJ}(2P)$, ISR, and \etab signals, the mass of the \etab, the inclusive background shape parameters, and an overall \egamma scale offset free to vary in the fit.
An \etab width of 10\mev is assumed.

Figure \ref{fig_y3s_high} shows the converted photon energy spectrum and fitted yields before and after the subtraction of the inclusive background, with an inset focusing on the \egamma region of the expected $\Upsilon(3S)\rightarrow\gamma\etab$ transition.
The results are summarized in Table \ref{tab_y3s_high}.
Although the $\chi_{b1,2}(1P)$-related peaks overlap, the \egamma resolution is still sufficient to measure the separate contributions.
We find no evidence for $\chi_{b0}(2P)\rightarrow\gamma\Upsilon(1S)$ decay.
The $\Upsilon(1S)$ yield from ISR production is in greement with the expectation from the previous {\babar\ }measurement \cite{babar_etab1}.
The best fit for a signal in the \egamma range corresponding to $\Upsilon(3S)\rightarrow\gamma\etab$ has $\egamma\approx908$\mev, which is a departure from, but not significantly inconsistent with, the nominal PDG value of $920.6^{+2.8}_{-3.2}$\mev \cite{pdg}.
Estimating the statistical significance from the change in $\chi^{2}$ of the fit with and without this component results in the equivalent of a less than $2.9\sigma$ effect.
Based predominantly on the positions of the $\chi_{b1,2}(2P)$ transition peaks, the \egamma scale offset in this energy range is $-0.9^{+0.4}_{-0.9}$\mev.
We further verify that the \egamma scale is correct by repeating the fit with the peak positions of the $\chi_{bJ}(2P)$ and ISR components allowed to vary, and they are found at the expected locations.
We also repeat the analysis with the \egamma scale offset forced to reproduce an \etab result corresponding to the \egamma value for the nominal $m_{\etab}$.
The assumption that the observed mass difference is due to an offset in the energy scale by $\sim12\mev$ is completely inconsistent with the photon energies observed for the well-established $\chi_{b1,2}(2P)$ states.
Even with only a $5\mev$ shift, the fit returns $\chi_{bJ}(2P)\rightarrow\gamma\Upsilon(1S)$ yields that disagree with the world average \cite{pdg} by more than a factor of two, and a $\chi^{2}/ndof\approx840/492$.

\begin{table*}
\begin{center}
\caption {Summary of the analysis of the $600 \leq E_{\gamma}^{*} \leq 1100$ \mev region of the $\Upsilon(3S)$ data. The \egamma column lists the transition energy assumed in this analysis, or in the case of $\Upsilon(3S)\rightarrow\gamma\etab$, the most significant $(\sim2.7\sigma)$ feature in the relevant \egamma region. Errors on the yield are statistical only. Regarding the derived branching fractions: the {\babar\ }value is from this paper, while the CUSB and CLEO columns are derivations based on \cite{cusb_chib} and \cite{cleo2} using up-to-date secondary branching fractions from \cite{pdg}. For the {\babar\ }result, the listed uncertainties are statistical, systematic, and from the uncertainties on secondary branching fractions, respectively. Upper limits are given at the $90\%$ confidence level. Dashes indicate no value has been measured in the quoted reference.}
\begin{tabular}{cccccccc}
\hline \hline
Transition & \egamma & Yield & $\epsilon$ & \multicolumn{3}{c}{Derived Branching Fraction $(\%)$}\\
& (\mev) & & ($\%$) & {\babar} & CUSB & CLEO \\
\hline
$\chi_{b0}(2P)\rightarrow\gamma\Upsilon(1S)$ & $742.7$ & $469^{+260}_{-259}$ & 1.025 & $0.7\pm0.4^{+0.2}_{-0.1}\pm0.1 \:(<1.2)$ & $<1.9$ & $<2.2$\\
$\chi_{b1}(2P)\rightarrow\gamma\Upsilon(1S)$ & $764.1$ & $14965^{+381}_{-383}$ & 1.039 & $9.9\pm0.3^{+0.5}_{-0.4}\pm0.9$ & $7.5\pm1.3$ & $10.4\pm2.4$\\
$\chi_{b2}(2P)\rightarrow\gamma\Upsilon(1S)$ & $776.4$ & $11283^{+384}_{-385}$ & 1.056 & $7.0\pm0.2\pm0.3\pm0.9$ & $6.1\pm1.2$ & $7.7\pm2.0$\\
$\Upsilon(3S)\rightarrow\gamma\etab$ & $907.9\pm2.8\pm0.9$ & $933^{+263}_{-262}$ & $1.388$ & $0.058\pm0.016^{+0.014}_{-0.016} \:(<0.085)$ & - & -\\
\hline \hline
\end{tabular}
\label{tab_y3s_high}
\end{center}
\end{table*}

The sources of systematic uncertainty and their evaluation are identical to those listed in Sec. \ref{sec_y2s_high}.
The main difference between the two energy regions is that, as previously remarked, the improved efficiency and background conditions in the $600 \leq E_{\gamma}^{*} \leq 1100$ \mev region of the $\Upsilon(3S)$ dataset lead to fit results that are more stable.
For the $\chi_{b1,2}(2P)$-related measurements, the dominant systematic uncertainty is due to the conversion efficiency ($3.6\%$), and all other sources are less than $2\%$.
For the \etab signal, the largest uncertainty in the yield is related to the assumed \etab width ($^{+17}_{-27}\%$).
Of the remaining systematic uncertainties, the largest two are due to the MC parameterization ($\sim15\%$) and bottomonium masses ($\sim4\%$), both enhancing the yield in a positive direction.
Uncertainty due to the background shape, the largest factor in the equivalent $\Upsilon(2S)$ analysis, is well controlled in the $\Upsilon(3S)$ dataset and contributes less than $3\%$ to the total uncertainty.
The uncertainty in \egamma is dominated by statistical uncertainty, and the largest systematic contribution is related to uncertainty in the \egamma scale via the uncertainty in the other bottomonium masses \cite{pdg}.

We measure $\mathcal{B}(\Upsilon(3S)\rightarrow\gamma\chi_{bJ}(2P))\times\mathcal{B}(\chi_{bJ}(2P)\rightarrow\gamma\Upsilon(1S))$ = $(3.9\pm2.2^{+1.2}_{-0.6})\times 10^{-4}$, $(12.4\pm0.3\pm0.6)\times 10^{-3}$, and $(9.2\pm0.3\pm0.4)\times 10^{-3}$, for $J=$ 0, 1 and 2, respectively.
Using $\mathcal{B}(\Upsilon(3S)\rightarrow\gamma\chi_{bJ}(2P))$ from the PDG \cite{pdg}, we derive $\mathcal{B}(\chi_{bJ}(2P)\rightarrow\gamma\Upsilon(1S))$ = $(0.7\pm0.4^{+0.2}_{-0.1}\pm0.1)\%$, $(9.9\pm0.3^{+0.5}_{-0.4}\pm0.9)\%$, and $(7.0\pm0.2\pm0.3\pm0.9)\%$, where the uncertainties are statistical, systematic, and from the uncertainty on $\mathcal{B}(\Upsilon(3S)\rightarrow\gamma\chi_{bJ}(2P))$, respectively.
From these values, we calculate a $90\%$ confidence level upper limit of $\mathcal{B}(\chi_{b0}(2P)\rightarrow\gamma\Upsilon(1S))<1.2\%$.
As before, we rescale the previous results \cite{cusb_chib,cleo2} using the relevant branching fractions \cite{pdg} to produce the values for comparison in Table \ref{tab_y3s_high}.
For the $\mathcal{B}(\chi_{b0}(2P)\rightarrow\gamma\Upsilon(1S))$ value from CUSB II \cite{cusb_chib}, we convert the result to an upper limit of $<1.9\%$ at the $90\%$ confidence level.
Our $\chi_{bJ}(2P)$ transition results agree with the previous measurements, and are the most precise measurements to date.
Assuming the peak near $\egamma=900$\mev to be due to decays to \etab, our best fit result is $\mathcal{B}(\Upsilon(3S)\rightarrow\gamma\etab)=(5.8\pm1.6^{+1.4}_{-1.6})\times 10^{-4}$.
The total significance of this result once systematic uncertainties are included is $\sim2.7\sigma$, and we set a limit of $\mathcal{B}(\Upsilon(3S)\rightarrow\gamma\eta_{b}(1S)<8.5\times 10^{-4}$.
We repeat the fit with the \etab mass constrained to the PDG value and its uncertainty \cite{pdg}.
The significance of this constrained result is $<1.9\sigma$.
We measure $\mathcal{B}(\Upsilon(3S)\rightarrow\gamma\eta_{b}(1S)=3.8\pm1.6^{+0.9}_{-1.0})$, which translates into an upper limit of $\mathcal{B}(\Upsilon(3S)\rightarrow\gamma\eta_{b}(1S)<6.1\times 10^{-4}$.

\section{DISCUSSION}\label{sec_conclusion}
To conclude, we review the results of this study and their broader implications.
The results for $\mathcal{B}(\chi_{bJ}(nP)\rightarrow\gamma\Upsilon(mS))$ presented here are the first derived directly from a measurement of the photon spectrum.
For $J=$ 1 and 2, we have made some of the most precise measurements of these branching fractions to date, thus helping to resolve some discrepancies between previous experimental results (\emph{i.e.} in $\chi_{bJ}(2P)\rightarrow\gamma\Upsilon(2S)$ decays).
Table \ref{tab_bf_results_vs_theory} shows a comparison of our results with some theoretical predictions \cite{y1d_kwong_rosner}.
These predictions are in reasonable agreement with our experimental results.

\begin{table}
\begin{center}
\caption{Comparison of the experimental branching fraction results from this work ({\babar}) and some theoretical predictions \cite{y1d_kwong_rosner}.}
\begin{tabular}{ccc}
\hline \hline
Decay & {\babar} $(\%)$ & Theory $(\%)$\\
\hline $\mathcal{B}(\chi_{b0}(2P)\rightarrow\gamma\Upsilon(2S))$ & $(<2.8)$ & 1.27 \\ 
$\mathcal{B}(\chi_{b1}(2P)\rightarrow\gamma\Upsilon(2S))$ & $18.9\pm2.4$ & 20.2\\
$\mathcal{B}(\chi_{b2}(2P)\rightarrow\gamma\Upsilon(2S))$ & $8.3\pm1.4$ & 10.1\\
\hline $\mathcal{B}(\chi_{b0}(2P)\rightarrow\gamma\Upsilon(1S))$ & $(<1.2)$ & 0.96 \\ 
$\mathcal{B}(\chi_{b1}(2P)\rightarrow\gamma\Upsilon(1S))$ & $9.9\pm1.1$ & 11.8\\
$\mathcal{B}(\chi_{b2}(2P)\rightarrow\gamma\Upsilon(1S))$ & $7.0\pm1.0$ & 5.3\\
\hline $\mathcal{B}(\chi_{b0}(1P)\rightarrow\gamma\Upsilon(1S))$ & $(<4.6)$ & 3.2 \\ 
$\mathcal{B}(\chi_{b1}(1P)\rightarrow\gamma\Upsilon(1S))$ & $34.9\pm3.1$ & 46.1\\
$\mathcal{B}(\chi_{b2}(1P)\rightarrow\gamma\Upsilon(1S))$ & $19.5^{+1.8}_{-1.9}$ & 22.2\\
\hline \hline
\end{tabular}
\label{tab_bf_results_vs_theory}
\end{center}
\end{table}

Our observations of $\Upsilon(3S)\rightarrow\gamma\chi_{b0,2}(1P)$ decays confirm the general features seen in previous measurements \cite{cleo_inclusive, cleo_hadrons, cleo_new}: decays to $J=1$ are suppressed compared to $J=$ 2 and 0.
This is unusual compared to all other $S\rightarrow P$ radiative transitions in the heavy quarkonium system measured thus far.
As noted previously \cite{theory_overlap}, the wavefunction overlap in the $\langle 3^{3}S_{1}\mid r\mid 1^{3}P_{J} \rangle$ matrix elements is unusually small.
Therefore, predictions for these decay rates are largely dependent on higher-order relativistic corrections and are thus sensitive to specific details of the chosen theoretical model.
That said, the comparison of our results with a selection of theoretical predictions \cite{theory_overlap, theory_list} shown in Table \ref{tab_s3_to_p1} (where we have converted our branching fraction measurements into partial widths) finds no good agreement with any particular model.
Indeed, even the hierarchy of the decay rates ($J=2>0>1$) is generally not well predicted.
Further work, both theoretical and experimental, will be required to understand these decays.

\begin{table}
\begin{center}
\caption{Comparison of our results with predictions \cite{theory_overlap, theory_list} for $\Upsilon(3S)\rightarrow\gamma\chi_{bJ}(1P)$ decays. We convert our result into partial widths (in units of eV) using a total width of $\Gamma_{\Upsilon(3S)}=20.32\pm1.85$ keV \cite{pdg}, absorbing this additional uncertainty into the total.}
\begin{tabular}{lccc}
\hline \hline
Source & $J=0$ & $J=1$ & $J=2$\\
\hline
{\babar} & $55\pm10$ & $<22$ & $216\pm25$\\
\hline
Moxhay-Rosner & $25$ & $25$ & $150$\\
Grotch \emph{et al.}  & $114$ & $3.4$ & $194$\\
Daghighian-Silverman & $16$ & $100$ & $650$\\
Fulcher & $10$ & $20$ & $30$\\
L\"{a}hde & $150$ & $110$ & $40$\\
Ebert \emph{et al.} & $27$ & $67$ & $97$\\
\hline \hline
\end{tabular}
\label{tab_s3_to_p1}
\end{center}
\end{table}

The searches for $\eta_{b}(1S)$ and $\eta_{b}(2S)$ states using the converted photon energy spectrum are largely inconclusive.
Over a range of approximately $9974<m_{\eta_{b}(2S)}<10015$\mevcc, we find $\mathcal{B}(\Upsilon(3S)\rightarrow\gamma\eta_{b}(2S))<1.9\times 10^{-3}$.
This value is consistent with, but does not improve upon, previous measurements \cite{cleo_inclusive}.
Due to low efficiency and high background, no evidence for $\Upsilon(2S)\rightarrow\gamma\etab$ is found.
In the $\Upsilon(3S)$ system, the most significant peaking structure in the \egamma energy region expected for the $\Upsilon(3S)\rightarrow\gamma\etab$ transition has $\egamma\approx908$\mev.
If interpreted as an \etab signal, this value trends toward the most recent potential model \cite{theory_etab_models} and lattice \cite{theory_etab_lattice} predictions, but we caution that the significance of this result is insufficient to draw such a conclusion regarding the \etab mass.
Taking advantage of the improved resolution from a converted photon technique to make a definitive measurement of the \etab mass and width will require more data from future experiments.

\section{ACKNOWLEDGEMENTS}\label{sec_acknowledgements}
\input pubboard/acknowledgements.tex

\appendix
\section{SYSTEMATIC UNCERTAINTIES ON MC-DETERMINED EFFICIENCIES}\label{appendix}
Branching fraction measurements in this analysis rely on MC-generated signal decays to determine the photon conversion and reconstruction efficiency.
This efficiency is dependent on the detector material model.
To evaluate a systematic effect due to the understanding of the detector in the simulation, a comparison of $\epem\gamma$ and $\mu^{+}\mu^{-}\gamma$ samples between data and MC is made.
Inclusive decays to an $\epem$ or $\mu^{+}\mu^{-}$ pair plus a photon are selected by requiring exactly four charged tracks in the event.
The CM momentum of the two highest-momentum non-conversion tracks as a fraction of half of the CM beam energy ($x_{1}^{*}$, $x_{2}^{*}$), the higher and lower values of their CM polar angles ($|\cos\theta^{*}_{1,2}|$), and the CM acolinearity ($\alpha^{*}$), are used as discriminating variables.
We require $\epem\gamma$ events to pass a predefined filter optimized to select Bhabha scattering events, and for the $\mu^{+}\mu^{-}\gamma$ events to fail this requirement. 
In cases of multiple candidates per event, the candidate with $m_{\ell^{+}\ell^{-}\gamma}$ closest to the CM beam energy is retained.
The values for the selection criteria variables are summarized in Table \ref{tab_lumi_bad_cuts}.

\begin{table}
\begin{center}
\caption{Selection criteria for the $\epem\gamma$ and $\mu^{+}\mu^{-}\gamma$ efficiency studies.}
\begin{tabular}{ccc}
\hline \hline Quantity & $\epem(\gamma)$ & $\mu^{+}\mu^{-}(\gamma)$\\
\hline nTRK & $=4$ & $=4$\\
$x_{1}^{*}$ & $>0.75$ & $>0.85$\\
$x_{2}^{*}$ & $>0.50$ & $>0.75$\\
Greater $|\cos\theta^{*}|$ & $<0.70$ & $<0.70$\\
Lesser $|\cos\theta^{*}|$ & $<0.65$ & $<0.65$\\
$\alpha^{*}$ ($^{\circ}$) & $<30$ & $<20$\\
\hline \hline
\end{tabular}
\label{tab_lumi_bad_cuts}
\end{center}
\end{table}

To avoid contamination from resonant decays (\emph{e.g.} $\chi_{bJ}(nP)\rightarrow\gamma\Upsilon(mS)(\ell^{+}\ell^{-})$, or $\Upsilon(nS)\rightarrow\ell^{+}\ell^{-}$ plus an extraneous photon), only the off-peak datasets are used for this study.
The $\epem(\gamma)$ MC sample uses the BHWIDE generator \cite{bhwide}, while the $\mu^{+}\mu^{-}\gamma$ MC sample is generated using the KK2f generator \cite{kk2f}.
The acceptance-based cross sections for these processes used in the MC generation are calculated separately from this analysis as part of standard luminosity measurements in {\babar}.

A systematic correction to the MC-determined efficiency is determined by comparing the number of events expected from the luminosity-weighted MC samples with the total number reconstructed in the data.
The uncertainty on this correction (dominantly statistical) is used as the systematic uncertainty in the efficiency due to the detector material model.
The four samples ($\epem\gamma$ and $\mu^{+}\mu^{-}\gamma$ in off-peak $\Upsilon(2S)$ and $\Upsilon(3S)$ data) are averaged to calculate this number, as is justified by verifying excellent data-to-MC agreement across all relevant $\cos\theta$, \egamma, and $\rho_{\gamma}$ ranges.
Integrated over all events, the ratio of the data and MC is $96.3\pm3.1\%$ when modeling the photons converted in the detector material.
This value is applied as a correction factor, with $3.3\%$ (when considering cross section uncertainties of about $0.8\%$) taken as an estimate for the systematic uncertainty in the efficiency.

The MC-based signal efficiencies are also dependent on assumptions regarding inclusive bottomonium decays.
The $nTRK$ requirements attempt to select multihadronic final states.
A difference in $nTRK$ distributions between simulation and data could lead to an error on the reconstruction efficiency.
To determine the size of this effect, the analysis is repeated with the requirements $nTRK$ greater than 5 or $nTRK$ greater than 6.
The largest change in the efficiency-corrected yields for the most significant transitions ($\chi_{b1,2}(1,2P)\rightarrow\gamma\Upsilon(1,2S)$) is found to be $1.0\%$.

Uncertainty in the modelling of the $\pi^{0}$ veto efficiency is tested in a similar manner, by repeating the analysis with the veto excluded and examining the change in the fit results for the most statistically significant transitions.
We rescale the MC-derived efficiency to equal half of the difference between the weighted average of the nominal and non-$\pi^{0}$ vetoed results, and introduce a systematic uncertainty large enough to cover this difference.
Because lower energy photons are more susceptible to the application of a $\pi^{0}$ veto, we find the differences to be energy-dependent and assign a different correction and uncertainty for each energy region.
The scale factors (uncertainties) range from $0.991$ $(0.9\%)$ for the $600<\egamma<1100$\mev range in $\Upsilon(3S)$ data to $0.963$ $(3.9\%)$ in the $\Upsilon(2S)$ data.

We combine these values to estimate a total systematic uncertainty on the efficiency of $3.6\%$ to $5.2\%$, depending on the transition.

\end{document}

%% file: pubboard/authors_jan2011.tex
%% author list as of 05-Jan-2011 (391 authors)
%
\author{J.~P.~Lees}
\author{V.~Poireau}
\author{E.~Prencipe}
\author{V.~Tisserand}
\affiliation{Laboratoire d'Annecy-le-Vieux de Physique des Particules (LAPP), Universit\'e de Savoie, CNRS/IN2P3,  F-74941 Annecy-Le-Vieux, France}
\author{J.~Garra~Tico}
\author{E.~Grauges}
\affiliation{Universitat de Barcelona, Facultat de Fisica, Departament ECM, E-08028 Barcelona, Spain }
\author{M.~Martinelli$^{ab}$}
\author{D.~A.~Milanes$^{a}$}
\author{A.~Palano$^{ab}$ }
\author{M.~Pappagallo$^{ab}$ }
\affiliation{INFN Sezione di Bari$^{a}$; Dipartimento di Fisica, Universit\`a di Bari$^{b}$, I-70126 Bari, Italy }
% ...
\collaboration{The \babar\ Collaboration}
\noaffiliation

%% file: pubboard/acknowledgements.tex
We are grateful for the 
extraordinary contributions of our \pep2\ colleagues in
achieving the excellent luminosity and machine conditions
that have made this work possible.
The success of this project also relies critically on the 
expertise and dedication of the computing organizations that 
support \babar.
The collaborating institutions wish to thank 
SLAC for its support and the kind hospitality extended to them. 
This work is supported by the
US Department of Energy
and National Science Foundation, the
Natural Sciences and Engineering Research Council (Canada),
the Commissariat \`a l'Energie Atomique and
Institut National de Physique Nucl\'eaire et de Physique des Particules
(France), the
Bundesministerium f\"ur Bildung und Forschung and
Deutsche Forschungsgemeinschaft
(Germany), the
Istituto Nazionale di Fisica Nucleare (Italy),
the Foundation for Fundamental Research on Matter (The Netherlands),
the Research Council of Norway, the
Ministry of Education and Science of the Russian Federation, 
Ministerio de Ciencia e Innovaci\'on (Spain), and the
Science and Technology Facilities Council (United Kingdom).
Individuals have received support from 
the Marie-Curie IEF program (European Union), the A. P. Sloan Foundation (USA) 
and the Binational Science Foundation (USA-Israel).

%% file: PRD.bbl
\begin{thebibliography}{99}

\bibitem{general_theory}
Recent comprehensive reviews include: E.~Eichten \emph{et al.}, Rev. Mod. Phys. \textbf{80}, 1161 (2008); N.~Brambilla \emph{et al.}, Eur. Phys. J. C \textbf{71}, 1534 (2011); and the many references therein.

\bibitem{nomenclature}
Throughout this paper, we employ the following convention: angular momentum variable $J$ encompasses 0, 1, and 2 (1, 2 and 3) when referring to $\chi_{bJ}(nP)$ ($\Upsilon(1D_{J})$) states, and $m$ ($n$) includes 2 and 3 (1 and 2) when referring to $\Upsilon(mS)$ ($\chi_{bJ}(nP)$) principal quantum number. Only allowed transitions are considered.

\bibitem{cb1}
R.~Nernst \emph{et al.} (Crystal Ball Collaboration), Phys. Rev. Lett. \textbf{54}, 2195 (1985).
\bibitem{cb2}
W.S.~Walk \emph{et al.} (Crystal Ball Collaboration), Phys. Rev. D \textbf{34}, 2611 (1986).

\bibitem{argus_conv}
H.~Albrecht \emph{et al.} (ARGUS Collaboration), Phys. Lett. B \textbf{160}, 331 (1985).

\bibitem{cusb1}
F.~Pauss \emph{et al.} (CUSB Collaboration), Phys. Lett. B \textbf{130}, 439 (1983).
\bibitem{cusb2}
C.~Klopfenstein \emph{et al.} (CUSB Collaboration), Phys. Rev. Lett. \textbf{51}, 160 (1983).
\bibitem{cusb3}
M.~Narain \emph{et al.} (CUSB-II Collaboration), Phys. Rev. Lett. \textbf{66}, 3113 (1991).
\bibitem{cusb4}
U.~Heintz \emph{et al.} (CUSB-II Collaboration), Phys. Rev. Lett. \textbf{66}, 1563 (1991).
\bibitem{cusb_chib}
U.~Heintz \emph{et al.} (CUSB-II Collaboration), Phys. Rev. D \textbf{46}, 1928 (1992).
 
\bibitem{cleo1}
R.~Morrison \emph{et al.} (CLEO Collaboration), Phys. Rev. Lett. \textbf{67}, 1696 (1991).
\bibitem{cleo2}
G.~Crawford \emph{et al.} (CLEO Collaboration), Phys. Lett. B \textbf{294}, 139 (1992).
\bibitem{cleo3}
K.W.~Edwards \emph{et al.} (CLEO Collaboration), Phys. Rev. D \textbf{59}, 032003 (1999).
\bibitem{cleo_inclusive}
M.~Artuso \emph{et al.} (CLEO Collaboration), Phys. Rev. Lett. \textbf{94}, 032001 (2005).
\bibitem{cleo_new}
M.~Kornicer \emph{et al.} (CLEO Collaboration), Phys. Rev. D \textbf{83}, 054003 (2011).
\bibitem{cleo_conv}
P.~Haas \emph{et al.} (CLEO Collaboration), Phys. Rev. Lett. \textbf{52}, 799 (1984).

\bibitem{babar_etab1}
B.~Aubert \emph{et al.} ({\babar\ }Collaboration), Phys. Rev. Lett. \textbf{101}, 071801 (2008).
\bibitem{babar_etab2}
B.~Aubert \emph{et al.} ({\babar\ }Collaboration), Phys. Rev. Lett. \textbf{103}, 161801 (2009).

\bibitem{cleo_etab}
G.~Bonvicini \emph{et al.} (CLEO Collaboration), Phys. Rev. D \textbf{81}, 031104(R) (2010).

\bibitem{babar_detector}
B.~Aubert \emph{et al.} ({\babar\ }Collaboration), Nucl. Instrum. Methods Phys. Res., Sect. A \textbf{479}, 1 (2002).
\bibitem{babar_lst}
W.~Menges, IEEE Nucl. Sci. Symp. Conf. Rec. \textbf{5}, 1470 (2006); M.R.~Convery \emph{et al.}, Nucl. Instrum. Methods Phys. Res., Sect. A \textbf{556}, 134 (2006).
\bibitem{semantics}
Throughout, we adopt the convention that unless otherwise indicated, a single quoted uncertainty is the total uncertainty, and for a pair of uncertainties, the first is statistical and the second systematic.
\bibitem{evtgen}
D.J.~Lange, Nucl. Instrum. Methods Phys. Res., Sect. A \textbf{462}, 152 (2001).
\bibitem{jetset}
T.~Sj\"{o}strand, Comput. Phys. Commun. \textbf{82}, 74 (1994).
\bibitem{helicity}
See for example, L.S.~Brown and R.N.~Cahn, Phys. Rev. D \textbf{13}, 1195 (1976); G.~Karl, S.~Meshkov, and J.L.~Rosner, Phys. Rev. D \textbf{13}, 1203 (1976).
\bibitem{geant4}
S.~Agostinelli \emph{et al.} (Geant4 Collaboration), Nucl. Instrum. Methods Phys. Res., Sect. A \textbf{506}, 250 (2003).

\bibitem{pdg}
K.~Nakamura \emph{et al.} (Particle Data Group (PDG)), J. Phys. G \textbf{37}, 075021 (2010).
\bibitem{thrust}
S.~Brandt \emph{et al.}, Phys. Lett. \textbf{12}, 57 (1964).
\bibitem{fox-wolfram}
G.C.~Fox and S.~Wolfram, Nucl. Phys. B \textbf{149}, 413 (1979).
\bibitem{crystal_ball}
M.J.~Oreglia, SLAC-R-236 (1980); J.E.~Gaiser, SLAC-R-255 (1982); T.~Skwarnicki, DESY-F31-86-02 (1986).

\bibitem{cleo_y1d}
G.~Bonvicini \emph{et al.} (CLEO Collaboration), Phys. Rev. D \textbf{70}, 032001 (2004).
\bibitem{babar_y1d}
P.~del~Amo~Sanchez \emph{et al.} ({\babar\ }Collaboration), Phys. Rev. D \textbf{82}, 111102(R) (2010).
\bibitem{y1d_godfrey_rosner}
S.~Godfrey and J.L.~Rosner, Phys. Rev. D \textbf{64}, 097501 (2001); ibid. \textbf{66}, 059902(E) (2002).
\bibitem{y1d_kwong_rosner}
W.~Kwong and J.L.~Rosner, Phys. Rev. D \textbf{38}, 279 (1988).
\bibitem{ul}
The upper limit (UL) is calculated from $\int^{UL}_{0}G(x) dx / \int^{+\infty}_{0}G(x) dx = 0.9$, where $G(x)$ is a Gaussian with mean equal to the central value of the branching fraction measurement and standard deviation equal to the total uncertainty. This procedure is used throughout.

\bibitem{cleo_hadrons}
D.M.~Asner \emph{et al.} (CLEO Collaboration), Phys. Rev. D \textbf{78}, 091103 (2008).
\bibitem{babar_pipihb}
J.P.~Lees et al. ({\babar\ }Collaboration), Phys. Rev. D \textbf{84}, 011104(R) (2011).
\bibitem{cleo_dipion_shape}
D.~Cronin-Hennessy \emph{et al.} (CLEO Collaboration), Phys. Rev. D \textbf{76}, 072001 (2007).
\bibitem{cleo_dipion}
C.~Cawlfield \emph{et al.} (CLEO Collaboration), Phys. Rev. D \textbf{73}, 012003 (2006).
\bibitem{theory_dipion}
T.-M.~Yan, Phys. Rev. D \textbf{22}, 1652 (1980); Y.-P.~Kuang and T.-M.~Yan, Phys. Rev. D \textbf{24}, 2874 (1981).
\bibitem{benayoun}
M.~Benayoun \emph{et al.}, Mod. Phys. Lett. A \textbf{14}, 2605 (1999).
\bibitem{hb_godfrey_rosner}
S.~Godfrey and J.L.~Rosner, Phys. Rev. D \textbf{66}, 014012 (2002).
\bibitem{babar_pi0hb}
J.P.~Lees \emph{et al.} ({\babar\ }Collaboration), arXiv:1102.4565, submitted to Phys. Rev. D(R) (2011).

\bibitem{theory_overlap}
See for example, P.~Moxhay and J.L.~Rosner, Phys. Rev. D \textbf{28}, 1132 (1983); H.~Grotch, D.A.~Owen, and K.J.~Sebastian, Phys. Rev. D \textbf{30}, 1924 (1984).
\bibitem{theory_list}
F.~Daghighian and D.~Silverman, Phys. Rev. D \textbf{36}, 3401 (1987); L.P.~Fulcher, Phys. Rev. D \textbf{42}, 2337 (1990); T.A.~L\"{a}hde, Nucl. Phys. A \textbf{714}, 183 (2003); D.~Ebert, R.N.~Faustov, and V.O.~Galkin, Phys. Rev. D \textbf{67}, 014027 (2003).
\bibitem{theory_etab_models}
S.~Recksiegel and Y.~Sumino, Phys. Lett. B \textbf{578}, 369 (2004); B.A.~Kniehl \emph{et al.}, Phys. Rev. Lett. \textbf{92}, 242001 (2004); ibid. \textbf{104}, 199901(E) (2010).
\bibitem{theory_etab_lattice}
A.~Gray \emph{et al.} (HPQCD and UKQCD Collaborations), Phys. Rev. D \textbf{72}, 094507 (2005); T.~Burch \emph{et al.} (Fermilab and MILC Collaborations), Phys. Rev. D \textbf{81}, 034508 (2010); S.~Meinel, Phys. Rev. D \textbf{82}, 114502 (2010).

\bibitem{bhwide}
S.~Jadach, W.~Placzek, and B.F.L.~Ward, Phys. Lett. B \textbf{390}, 298 (1997).
\bibitem{kk2f}
S.~Jadach, B.F.L.~Ward, and Z.~Was, Comput. Phys. Commun. \textbf{130}, 260 (2000).


\end{thebibliography}
